\documentclass{aa}  
\usepackage{bm}
\usepackage{graphicx}
\usepackage{txfonts}

\usepackage[dvipsnames]{xcolor}
\usepackage{ulem}
\usepackage{color,soul}
\usepackage[colorlinks=true,linkcolor=blue,citecolor=blue,urlcolor=blue]{hyperref}
\newcommand{\yr}{\ensuremath{\mathrm{yr}}}

\newcommand{\B}{\ensuremath{\left|B\right|} }

\newcommand{\mdot}{\ensuremath{\dot{M}}}
\newcommand{\Msun}{\ensuremath{\,\mathrm{M}_{\odot}}}
\newcommand{\pyr}{\ensuremath{\,\mathrm{yr}^{-1}}}	
\newcommand{\vinf}{\ensuremath{\varv_{\infty}}}
\newcommand{\kms}{\ensuremath{\,~\textrm{km\,s}^{-1}}}	
\newcommand{\teff}{\ensuremath{T_{\mathrm{eff}}}}
\newcommand{\vrot}{\ensuremath{\varv_{\mathrm{rot}}}}
\newcommand{\vesc}{\ensuremath{\varv_{\mathrm{esc}}}}

\usepackage{ifthen}
\usepackage{xcolor}

\begin{document}

   \title{Multi-Dimensional MHD simulations of young Core-Collapse Supernova Remnants}

  % \subtitle{I. Overviewing the $\kappa$-mechanism}

   \author{C. J. K. Larkin
          \inst{1,2,3,4}
          \and
          J. Mackey\inst{5}
          \and
          B. Reville \inst{1}
          \and
          H. Jin \inst{6}
           \and
          N. Langer \inst{6}
          \and
          A. A. C. Sander \inst{2,7}}

   \institute{Max-Planck-Institut f\"{u}r Kernphysik, Saupfercheckweg 1, D-69117 Heidelberg, Germany\\
            \email{cormac.larkin@mpi-hd.mpg.de}
         \and Astronomisches Rechen-Institut, Zentrum f\"{u}r Astronomie der Universit\"{a}t Heidelberg, M\"{o}nchhofstr. 12-14, D-69120 Heidelberg, Germany \
         \and European Southern Observatory, Karl-Schwarzschild-Strasse 2, 85748 Garching bei M\"unchen, Germany \
         \and Max-Planck-Institut f\"{u}r Astronomie, K\"{o}nigstuhl 17, D-69117 Heidelberg, Germany \
         \and Astronomy \& Astrophysics Section, School of Cosmic Physics, Dublin Institute for Advanced Studies, DIAS Dunsink Observatory, Dublin D15 XR2R, Ireland \
         \and Argelander Institut f\"{u}r Astronomie, Auf dem H\"{u}gel 71, DE-53121 Bonn, Germany\
         \and Universit\"at Heidelberg, Interdisziplin\"ares Zentrum f\"ur Wissenschaftliches Rechnen, 69120 Heidelberg, Germany\label{inst:iwr}
             }
             % \thanks{The university of heaven temporarily does not
             %         accept e-mails}

   \date{Received September 15, 1996; accepted March 16, 1997}

% \abstract{}{}{}{}{} 
% 5 {} token are mandatory
 
  \abstract
  % context heading (optional)
  %{} %leave it empty if necessary  
   {Supernova remnants (SNRs) play a central role in shaping the structure and evolution of the interstellar medium. Core-Collapse Supernova (CCSN) progenitors are massive stars, which produce a dense circumstellar medium (CSM) through intense mass loss in the post-main sequence stages of stellar evolution. The subsequent supernova produces a strong shock which expands into a highly structured, complex magnetised environment.}
  % aims heading (mandatory)
   {Magnetohydrodynamic (MHD) consideration of pre- and post-SN evolution in multi-D are desirable to further our understanding of non-thermal aspects. We aim to determine how detailed stellar evolution treatment influences the shock propagation. We focus on two prototypical CCSN scenarios: Red Supergiants (RSGs), with slow stellar winds and moderate mass-loss rates, and Wolf-Rayet (WR) stars, with faster winds and higher mass-loss rates.}
  % methods heading (mandatory)
   {We use the \textsc{pion} code to perform 3D MHD simulations of these CCSN progenitor scenarios. We use a detailed stellar evolution prescription to accurately and self-consistently model the pre-SN CSM and initialise supernova explosions to investigate the surrounding plasma environment.}
  % results heading (mandatory)
   {Our 2D and 3D treatment, inclusion of radiative cooling and assumption of full photoionization produces CSM features that have not been identified in previous work. In the WR model we produce a coherent set of fast reflected shocks. In both cases we find faster forward shocks than predicted by analytic theory due to additional wind acceleration from photoionization for the RSG case, and accounting for the CSM expansion in the WR case. Model predictions of slowly rotating RSG and WR stars results in weakly magnetised wind bubbles, limiting potential for their SNRs to become PeV particle accelerators.}
  % conclusions heading (optional), leave it empty if necessary 
   {Detailed multi-D MHD treatment of the CSM is needed to account for SNR evolution beyond the wind termination shock, where dynamic instabilities can be important. Including self-consistent stellar evolution is important for determining the CSM density and magnetic field structure close to the star, which determines the shock properties and SNR evolution for the first few hundred years.}
   \maketitle
\nolinenumbers

%
%-------------------------------------------------------------------

\section{Introduction}

Young Supernova Remnants (SNRs) are of considerable interest to the high-energy astrophysics community, as they remain the most plausible sources for the production of cosmic rays (CRs) in our Galaxy \citep{GinzburgBook}.
While the arguments supporting an SNR origin for Galactic CRs have gained observational backing, especially at GeV to TeV energies \citep[e.g.,][]{FermiSNRs, HESS_SNR}, many questions remain unresolved.
A critical issue for the CR origins theory is the ``knee'' feature observed in the local CR spectrum at a few peta-electronvolts (PeV).
For an individual SNR to accelerate protons to such energies, a supernova (SN) must explode in environments with favourable conditions \citep{BELL2013, Vieu22, BROSE2025}.
The initial decades of an SNR's evolution, when the shock's energy processing rate ($\propto \rho \varv_{\rm sh}^3$) is at its peak, are believed to be crucial. During this time the SNR shock is expanding into the circumstellar medium (CSM) that has been pre-shaped by the progenitor's stellar wind as it evolved towards core-collapse.

Massive stars are copious sources of ionising radiation and drive powerful stellar winds throughout their lifetime \citep{Langer2012}, which determines the CSM within a radius of at least a few parsecs by the time of core-collapse \citep{GarciaSegura1996a_RSG_WR, GarciaSegura1996b_LBV, FICHTNER2024}, which may lead to a core-collapse supernova (CCSN).
A freely expanding stellar wind generates a bubble with a density profile $\rho\propto 1/r^2$, culminating in a wind termination shock (WTS) with size scale determined by the confining external ISM pressure \citep{Dyson1972}.
As massive stars evolve, they transform into various post-main sequence objects, such as Red Supergiants (RSGs), Wolf-Rayet stars (WRs) and Luminous Blue Variables (LBVs).
This evolution is marked by changes in their mass-loss rates (\mdot) and terminal wind velocities (\vinf), by as much as 2-3 orders of magnitude over timescales comparable to the wind advection timescale.
Consequently, a complex circumstellar medium (CSM) may emerge, comprising dense shells, rarefied bubbles and multiple shocks.

Hydrodynamic (HD) simulations of circumstellar nebulae by \citet{GarciaSegura1996a_RSG_WR, GarciaSegura1996b_LBV} found that shells from different stellar evolution phases interact and exhibit dynamical instabilities, underscoring the necessity for multi-dimensional simulations. More recent studies show that 3D simulations obtain stronger instability development than 2D because of the extra degree of freedom \citep{vanMarle2012}.
Radiative cooling can also influence the resulting CSM as it causes material to condense downstream of a shock, leading to larger downstream-to-upstream density ratios compared to standard hydrodynamic Rankine-Hugoniot shock jump conditions \citep{Shu1992}.

Simulations of CCSNe in circumstellar bubbles \citep{TenorioTagle1990,DWARKADAS2005,DWARKADAS2007} demonstrated that, as expected, the early evolution of the supernova remnant (SNR) is influenced by the ratio of ejecta mass to CSM mass and, consequently, the progenitor's mass-loss history.
The impact of the WR phase was examined by \cite{vanVeelen2009}, who found that the presence and duration of the WR phase affect the velocities of the reverse shock and shocked material. Other pathways to stripped-progenitor CCSNe via binary interactions were explored by \citet{Yasuda2021,Yasuda2022} and \citet{Ercolino2024a,Ercolino2024b}. Motion of the progenitor star through the ISM can lead to asymmetric CSM \citep{Meyer2015,Meyer2017}, resulting in pronounced asymmetries in the resulting SNR as it propagates through over-dense (or rarefied) regions in (or opposite to) the direction of stellar motion.
The supernova explosion itself is also expected to be asymmetric, based on both simulations \citep{Wongwathanarat2017} and observations of young SNRs such as 1987A \citep{Boggs2015} and Cas A \citep{Milisavljevic2013, Wang2016}.
Simulations that combine an asymmetric explosion model with a detailed CSM have been conducted for several young SNRs \citep[e.g.,][]{Orlando2021,Orlando2025GM,Orlando20251987}.

These literature results demonstrate that accurate models of the external density, velocity ($\mathbf{V}$), and magnetic field ($\bm{B}$) profiles are important for early SNR modelling.
Additionally, the stellar rotation velocity, \vrot, which winds up the magnetic field, sets up the magnetic orientation of the SNR shock.
The surface magnetic field strength, $B_\star$, is key for setting the $\bm{B}$ profile close to the star, while at parsec scales and above, interstellar fields have increasing importance \citep{vanMarle2015}.
Downstream of a spherical WTS in the subsonic flow, the azimuthal magnetic field can increases as $B_{\phi} \propto r$ as the flow compresses, the so-called Axford-Cranfill effect \citep{Axford1972,Cranfill74}.

These effects are influenced by the non-uniformity of the ISM, stellar proper motion, and instabilities.
To explore this, simulations have been conducted to assess the particle acceleration potential for different supernova progenitors, such as RSGs, WRs, and LBVs.
Much of this research has been performed in 1D, utilising simplified stellar evolution models where the circumstellar medium (CSM) from only one evolutionary phase is considered, and average CSM densities are assumed \citep{Telezhinsky2012b,Telezhinsky2013,BROSE2022}.
These studies often involve assumptions about the magnetic field configuration, as developing a comprehensive magnetic field model typically requires at least 2D simulations. 

\citet{ZP2018} performed a 2D MHD simulation of the expansion of a WR wind from a runaway star into a uniform ISM, with constant wind properties $\dot{M} =10^{-5}$\,\Msun\,\text{yr}$^{-1}$ and $\vinf =1000$\,\kms. Assuming a stellar radius $R_{\star}=10^{12}$ cm, we can infer surface magnetic field $B_\star=125$\,G and rotation $\vrot=100\,\kms$. The study found that the Axford-Cranfill effect led to significant magnetic field accumulation downstream of the WTS, which leads to favourable conditions for particle acceleration to PeV energies in the SNR blastwave.

Effects such as curvature and gradient drifts cannot be fully accounted for using 1/2D simulations as particle acceleration is fundamentally a 3D process.
These effects are expected to be most relevant for the highest-energy particles, as discussed by \cite{Bell2008,ZP2018}, and thus may be key for determining the viability of CCSNe to achieve PeV energies.

Most of the aforementioned work was focused on the earliest times, when the SN expands into the freely-expanding wind region, and did not account for other sites of acceleration.
When the forward shock interacts with CSM structure outside of this region, additional reflected shocks can be produced.
A reflected shock can then interact with the SN reverse shock, as inferred in e.g. SNR 1987A \citep{Zhekov2009}, G330.2+1.0 \citep{Borkowski2018} and Cas A \citep{Vink2022,Fesen2025} and considered by \cite{Sushch2024}.
At later times, a reflected shock could propagate back into the low-density cavity evacuated by the SN explosion \citep{DWARKADAS2007, Meyer2015}.

To assess the potential of a SNR to accelerate particles, we seek an accurate description of two distinct phases. Firstly, we require detailed modelling of the pre-SN CSM in multi-D, accounting for detailed stellar evolution through multiple phases, accounting for dynamical instabilities and the effects of radiative cooling. Secondly, we then need to trace the evolution of a SN explosion through this CSM in 3D MHD.

In this paper, we focus on the pre-SN circumstellar environment and early stages of the SNR evolution.
The  goal is to qualitatively examine the resulting environmental and shock conditions that establish the particle acceleration potential of young core-collapse SNRs.
Our MHD treatment permits modelling of the average macroscopic conditions upstream of the SN forward shock in three dimensions.
Our CSM is generated from a stellar evolution model with time-dependent mass-loss.
We consider the two most common canonical progenitor scenarios, an exploding RSG and an exploding WR star. 

Our paper is organised as follows:
in Sect.~\ref{sec:method} we discuss how we implement stellar evolution (Sect.~\ref{subsec:stellar_models}), our MHD simulation setup (Sect.~\ref{subsec:mhd_sims}), details of our RSG (Sect.~\ref{subsec:RSG_sim}) and WR (Sect.~\ref{subsec:WR_sim}) CSM simulations and how we insert a SN (Sect.~\ref{subsec:SN_implementation}).
In Sect.~\ref{sec:results} we discuss the hydrodynamic and magnetic field evolution of the pre-SN CSM (Sect.~\ref{subsec:pre_results}), RSG (Sect.~\ref{subsec:RSG_results}) and WR (Sect.~\ref{subsec:WR_results}) simulations, and the properties of the forward shock in both cases (Sect.~\ref{subsec:shock_results}).
In Sect.~\ref{sec:discussion} we discuss the impacts of stellar evolution (Sect.~\ref{subsec:discussion_stevol}), stellar environment (Sect.~\ref{subsec:discussion_environment}), SNR evolution (Sect.~\ref{subsec:discussion_post_SN_evol}) and implications for our understanding of particle acceleration from young SNRs (Sect.~\ref{subsec:discussion_PA}).
We present our conclusions in Sect.~\ref{sec:conclusion}.

%--------------------------------------------------------------------
\section{Methods}
\label{sec:method}
\subsection{Stellar model}
\label{subsec:stellar_models}

In this work, we use a stellar evolutionary track computed with the 1D stellar evolution code \textsc{MESA} \citep[][r10398]{MESA_I, MESA_II, MESA_III, MESA_IV, MESA_V}. The evolutionary track is taken from the dense binary evolution model grid of \citet{Jin2025} for solar metallicity and corresponds to the 31.6$\,M_\odot$ primary star model from a wide binary system. This primary star has no interaction with the secondary during its evolution and so evolves as a single star. Here we briefly describe the relevant physics assumptions.

\begin{figure}[!tbp]
    \centering
    \includegraphics[width=8.8cm]{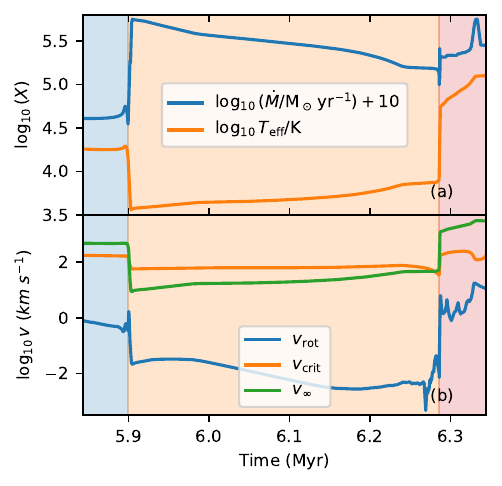}
    \caption{Stellar parameters for the last 500\,kyr of our evolutionary track. The blue, orange and red shading denote the main sequence, RSG and WR phases respectively. (a) Mass loss rate and surface temperature vs time since ZAMS. (b) Rotation velocity, critical rotation velocity and wind velocity vs time since ZAMS.}
    \label{fig:stevol}
\end{figure}

The mass-loss prescription relevant for our work uses the mass-loss rate of \citet{Vink2001} during the main sequence phase, that of \citet{Nieuwenhuijzen1990} for the RSG phase, and those of \citet{Nugis2000} and \citet{Yoon2017} for the WR phase. We show the time evolution of key stellar parameters for these phases in Fig. \ref{fig:stevol}. The star has an initial rotation of 20\% critical rotation at zero-age main sequence (ZAMS), which is average for Galactic O stars \citep{Holgado2022}. The evolutionary track has a main-sequence (MS) phase lasting 5.9\,Myr, a post-main-sequence RSG phase of $\sim$$400$\,kyr followed by a WR phase of $\sim$$50$\,kyr. The calculation is terminated at core helium depletion with a CO-core mass of $\sim$$11 M_\odot$, which may lead to explosion as a CCSN \cite[see, e.g.,][]{Aguilera-Dena2023}.

Our evolutionary track has \vrot\ $ = 11$\,\kms\, at the time of explosion, which corresponds well with model predictions for WR stars \citep[e.g.][]{Meynet2005}. Already by the end of the main sequence the initial $\varv_{\mathrm{rot}}$ has reduced from 200\,\kms\ to $<1$\,\kms\ because of expansion and mass loss (with consequent loss of angular momentum).
This further decreases to $\varv_{\mathrm{rot}}\sim 10^{-2}$\,\kms\ during the RSG phase as the star has expanded dramatically.
The increase at the WR phase arises because the outer layers are expelled in stellar wind and the more rapidly rotating stellar core is exposed, but still
the ratio $\varv_{\mathrm{rot}}/\varv_{\mathrm{crit}}$ is only of order 1\%.
The lack of rapid rotation has significant consequences for the evolution of the circumstellar magnetic field. It cannot be assumed to be fully tangential (\B$\sim B_{\phi}\propto 1/r$), but instead will be radially dominated up to a point $r_t$, defined as the radius at which $B_r / B_{\phi} = 1$ close to the equatorial plane. For $r \leq r_t$, \B$\sim B_r \propto 1/r^2$.  

A large uncertainty in performing simulations of this type is the choice of surface magnetic field strength $B_\star$ of the progenitor. Approximately 6-7\% of OB stars have detectable magnetic fields \citep{Grunhut2017,Scholler2017}, which can, for example, be generated in mergers \citep{Schneider2019,Frost2024}. For RSGs, a small number of surface magnetic field measurements  have been made \citep{Auriere2010,Tessore2017}, supporting a range of 1-10 G. The situation for WR stars is less clear. The fraction of WRs with strong fields is poorly constrained observationally due to the difficulties in making high-resolution spectropolarimetric observations of fast Doppler-shifted WR winds \citep{delaChevrotiere2014, Hubrig2020,Shenar2023}. Additionally, a strong surface field would inhibit mass-loss and thus reduce the density of the CSM \citep[e.g.,][]{Owocki2004,Frost2024}. Since $B_\star$ is not predicted by the \textsc{MESA} calculation, we choose a relatively weak (and constant in time) value of $B_\star=10$\,G at all evolutionary phases.

\begin{figure}[!htbp]
    \centering
    \includegraphics[width=1\linewidth]{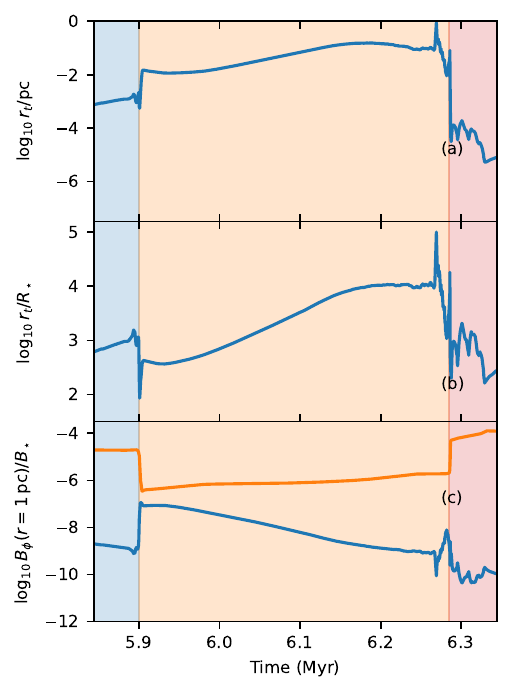}
    \caption{Evolution of selected CSM magnetic field components for the last 500\,kyr pre-SN. The blue, orange and red shading denote the main sequence, RSG and WR phases respectively. (a) $r_t$, the radius where $B_r / B_{\phi} = 1$ close to the equatorial plane, vs.~time. (b) $r_t/R_{\star}$ vs time. (c) $B_{\phi}/B_{\star}$ vs.~time where $B_\phi$ is calculated at a distance of 1\,pc from the star, for our model (blue) and with values of \citet{ZP2018} (orange). Note while $R_\star$ is time dependent, $B_\star$ is assumed to be fixed.}
    
    \label{fig:Bevol}
\end{figure}

In Fig. \ref{fig:Bevol} we show the evolution of selected magnetic field components in the pre-SN CSM using the analytical relation $r_t = R_{\star}\vinf/\vrot$, based on Eq.\,9.12 of \citet{LamersCassinelli1999}. For the RSG phase, $r_t\in [0.01,0.1]$\,pc, or $(10^3-10^4)\,R_\star$, with a predominantly radial magnetic field within this region. Consequently, the CSM magnetic field at pc scales is $10^{-7}-10^{-9}$ of the surface field. During the WR phase, the stellar radius is $\sim 10^3$ times smaller than during the RSG phase, and so the CSM field at pc scales is similar to that at the end of RSG phase (or even weaker), even though the surface rotation rate is significantly larger.

We use a single evolutionary track as the basis for both the RSG and WR simulations. We use the full evolutionary track for the WR simulation and use only the RSG phase for the RSG simulation. Using the same RSG phase for both models is similar to the approach of \cite{vanVeelen2009} and allows us to focus on the effect of the WR phase sweeping out the RSG material. This approach also ensures the self-consistency of our assumed stellar evolution. The two scenarios presented are chosen to qualitatively compare the slow, dense RSG CSM with the fast, rareified and highly structured WR CSM. These scenarios will not be representative for all CCSNe, but serve as exemplary cases to highlight some effects of detailed stellar evolution on the CSM and subsequent SNR evolution.

\subsection{MHD simulations}
\label{subsec:mhd_sims}
In this work, we perform 2D cylindrical and 3D Cartesian simulations using the MHD code \textsc{pion} \citep{2021PION} for our models. \textsc{pion} includes static mesh refinement, such that high resolution can be applied in regions of interest. The entire domain $D$ is included for refinement level $n=0$, and each successive level covers $D/2^n$. We use the evolving stellar wind source term introduced in \cite{2021PION}. 

The stellar evolution track described in \ref{subsec:stellar_models} is used to obtain the star's mass, luminosity, \teff, \mdot, \vrot, \vesc\ and radius as a function of time. These parameters determine the properties of our evolving stellar wind source. Using these values as our internal boundary conditions, the system is evolved, adopting ideal MHD, as described in \cite{2021PION} for the pre-SN evolution. The divergence cleaning method of \cite{DEDNER2002} is used throughout. The stellar magnetic field is injected as a split monopole that is swept into a spiral pattern by stellar rotation, following \citet{2021PION}.

Radiative heating and cooling is included following the scheme described in \cite{Green2019}. We assume photoionized gas at all times, for which the heating rate is dominated by the photoionization of recombining H$^+$ ions. We therefore calculate the heating as the recombination rate multiplied by a mean heating energy per ionization, assumed to be 5 eV as an average value for an O-type star \citep[cf.][]{Green2019}. This would not be strictly true for an isolated RSG, but in a cluster environment nearby O stars would photoionize a RSG's wind to within 0.05 pc of the star \citep{Mackey2015,Larkin25}.

Our radiative cooling prescription (cooling model 8 in \textsc{pion}) includes three cooling sources, as described in \citet{Green2019}. Firstly, we take the maximum of the collisional ionization equilibrium (CIE) curve for Solar metallicity described by \cite{Wiersma2009} and the forbidden-line cooling function described by \cite{Henney2009}, to capture the large cooling rates in photoionization equilibrium at $T\sim10^4$\,K. Secondly, we include Bremsstrahlung from ionized hydrogen \citep{Hummer1994} and helium \citep{Rybicki1979}. Finally, we use the hydrogen recombination cooling rate from \cite{Hummer1994}, assuming hydrogen is fully ionized. This heating and cooling prescription produces an equilibrium gas temperature $T\approx8300$\,K.

We compute the RSG and WR models with different setups. For the RSG model we evolve it fully in 3D starting from the end of the MS stage of the evolutionary track. This avoids the need to wrap the CSM from 2D. For the WR model, CSM from the previous stages is important. Given the computational constraints in evolving fast stellar winds at high spatial resolution, we evolve the simulation in 2D from 2\,Myr post ZAMS through the RSG stage to the end of core helium burning in 2D, and then wrap the simulation to 3D. This is achieved using a first-order interpolation, such that the coordinates of each 3D Cartesian cell are mapped onto the 2D cylindrical grid and the scalar quantities are copied directly from the 2D grid cell into the 3D grid cell.  Vector quantities from 2D such as $\mathbf{V}\equiv(\varv_z,\varv_R,\varv_\phi)$ are transformed to Cartesian coordinates $(\varv_x,\varv_y,\varv_z)$ assuming axisymmetry.

\subsection{RSG Model}
\label{subsec:RSG_sim}
The evolution is performed using a 3D cubic Cartesian domain of $(x,y,z) \in [-2.048\times 10^{20}, 2.048\times 10^{20}]$ cm. The 10 levels of static mesh-refinement focussed on the star at the origin give us a finest grid resolution of $\Delta x = 3.12\times10^{15}$ cm, or about 0.001 pc.
Each refinement level has $256^3$ grid cells, and the factor of two refinement per level means that successive levels have domains $2\times$ smaller along each dimension than the next coarser level.
The finest level has a cubic domain $(x,y,z) \in [-4\times10^{17},4\times10^{17}]$\,cm.
We choose an ambient magnetic field of $(B_x,B_y,B_z) = (4\times10^{-7},1\times10^{-7},1\times10^{-7})$ G.
To mimic proper motion of the star, we include a small velocity for the ISM, such that in the star's reference frame it is moving through the ISM with $(\varv_x,\varv_y,\varv_z) = (-4,1,4)$ \kms.
As the RSG wind would already be supersonic at rest, this small motion does not cause the wind to become supersonic, and the motion is included in order to break possible artificial symmetries that may arise from a fully stationary simulation.
Inflow boundaries are set at the outer boundaries with an inflow in the initial conditions, and zero-gradient (outflow) boundaries are applied otherwise.

\subsection{WR Model}
\label{subsec:WR_sim}
The WR simulation is evolved in a 2D cylindrical coordinate system $(r,z)$ (with assumed rotational symmetry around the $z$ axis in the angular $\phi$ coordinate), using a rectangular domain of $r\in[0, 2.048\times 10^{20}]$ cm, $z\in[-2.048\times 10^{20}, 2.048\times 10^{20}]$ cm. Ten levels of static mesh-refinement are applied, centred on the star at the origin, with a factor of two refinement between each level. Every level has $128\times 256$ grid cells, and so successive levels have domain ranges $2 \times$ smaller in each dimension. The finest level has a domain $r\in[0,4\times10^{17}]$\,cm and $z\in[-4\times10^{17},4\times10^{17}]$\,cm and a cell diameter $\Delta x = 3.12\times10^{15}$\,cm. Axisymmetric reflecting boundary conditions are applied at $r=0$, and zero-gradient (outflow) conditions imposed at the outer edges of the domain. We choose a wind radius of $r = 1.2\times10^{17}$ cm, corresponding to $38$ grid cells at the highest refinement level. We place the star at the origin in a uniform ISM. We choose an initial constant ambient ISM density of $\rho_0 = 2.338\times10^{-23}$ g cm$^{-3}$, corresponding to 10 hydrogen atoms per cm$^{3}$. We choose an ambient ISM pressure of $P_0 = 6.072\times10^{-12}$ dyne\,cm$^{-3}$, corresponding to a temperature of 4000 K.
We choose an ambient magnetic field of $B_z = 4\times10^{-7}$ G. We impose $B_r=0$ to avoid monopole generation on the $z$-axis, and $B_\phi=0$ because a large-scale circular magnetic field centred on the trajectory of the star is a very unlikely configuration.
The 2D axisymmetric simulation does not allow a more general magnetic field configuration that is misaligned with the grid axes, as was used above for the RSG model.
We do not include proper motion as the 2D to 3D wrapping requires assuming axisymmetry, and the effects of a small proper motion would be much less pronounced with the fast WR wind. 

\subsection{Supernova implementation}
\label{subsec:SN_implementation}
 To introduce a supernova into the 3D simulation, a spherical region is overwritten using an iterative two-component density and velocity profile following how \cite{WHALEN2008} and  \citet{FICHTNER2024} implemented the solution of \cite{TrueloveMcKee1999}. This profile is flat between $0 \leq r_\mathrm{core}$ and decreases sharply as a power law outside $r_\mathrm{core}$ to the maximum radius $r_{max}$. The power-law index we adopt is $n=9$, in line with previous CCSNe simulations \citep[e.g.,][]{TrueloveMcKee1999,DWARKADAS2007}. We choose an ejecta mass of $5M_\odot$ and energy of $10^{51}$\,erg \citep{Burrows2021}, giving a core velocity of $4.695\times10^{8}$ cm\,s$^{-1}$. We chose $r_{max} = 40$ cells $(\sim 0.04$ pc, corresponding to $t_{max}\sim 1.3$\,yr post-explosion). This is necessary to avoid grid artefacts in the supernova remnant as it expands. Furthermore, we impose Gaussian perturbations of $\pm 40\%$ for each cell of the SN, to approximate an aspherical explosion. This overwriting procedure results in a small discontinuity at the leading edge of the SN which is found to smooth out within the first output timestep.

We choose a split monopole profile for our ejecta magnetic field. This decision is primarily motivated by requiring the ejecta field lines to smoothly connect to the lines already in the CSM, and our inability to simulate a convective core that would produce a physically motivated magnetic field \citep{Varma2021}. We set the magnetic field strength such that the plasma $\beta$ is equal to the $\beta$ of the CSM ($\sim$$100$). 

\section{Results}
\label{sec:results}
We present results for our pre-SN CSM and SNR simulations, aiming to highlight differences between the RSG and WR cases arising from their different CSM. We present radial CSM profiles for both 3D SNR simulations. For the SNR simulations, we present XZ-plane slices for the density, the (normalised) compression \(-\nabla\cdot \mathbf{V} / {\rm V}\) and magnetic field strength. 

\subsection{Pre-SN CSM}
\label{subsec:pre_results}

\begin{figure}[!htbp]
    \centering
    \includegraphics[width=8.8cm]{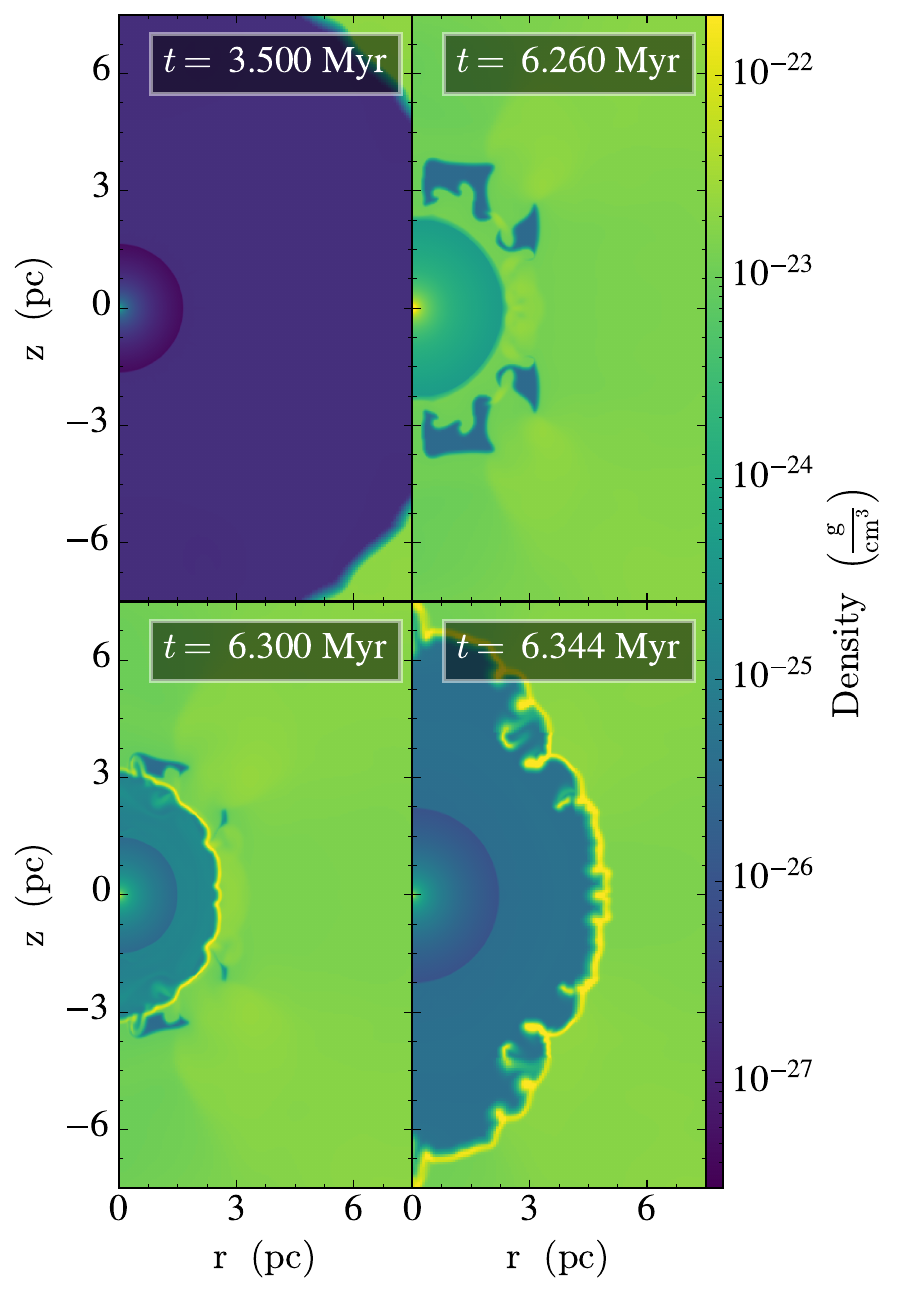}
    \caption{Density slices from the 2D evolution showing the CSM during the Main Sequence (upper left), RSG phase (upper right), WR phase sweeping out the RSG shell (lower left) and pre-SN (lower right).}
    \label{fig:2D_bubble}
\end{figure}

We show key evolutionary phases of the 2D CSM in Fig.~\ref{fig:2D_bubble}. The 2D and 3D CSM at the end of the RSG phase are quantitatively similar, with a large density jump at the edge of the RSG wind bubble and no MS wind bubble due to photoionization. The ISM magnetic field is assumed to be uniform and parallel to the $z$-axis, and therefore spherical expansion of the wind bubble induces magnetic tension that provides a restoring force in the directions perpendicular to $\bm{B}$ but not in the parallel direction. This was explored in detail by \citet{vanMarle2015}, who showed that the generic solution is a bubble elongated along the axis parallel to the ISM magnetic field, and with rotational symmetry about this axis.  We obtain a very similar result. 

\begin{figure*}[!htbp]
\centering
   \includegraphics[width=18cm]{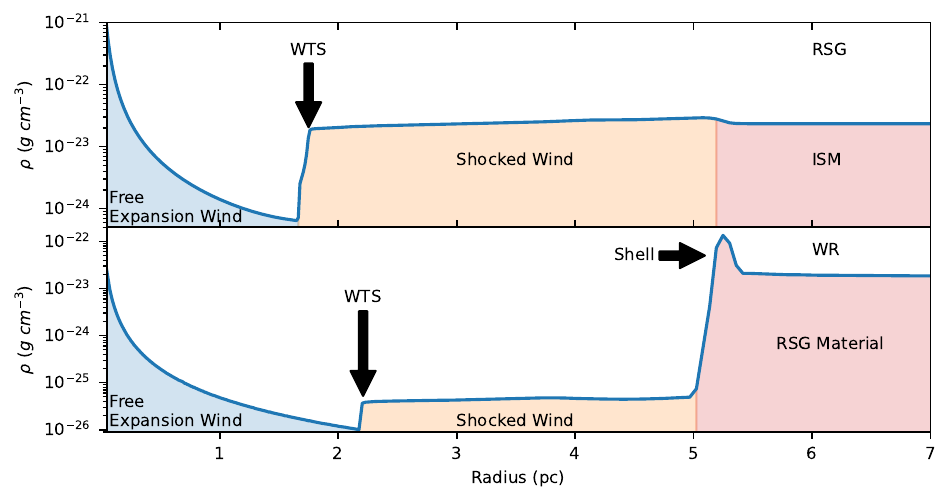}
     \caption{Radial profile of density for the RSG simulation (upper panel) and WR simulation (lower panel) before the SN is inserted. The profile is calculated along a diagonal ray from the origin in the $[+x,-y,+z]$ direction. 
     }
     \label{fig:pre_CSM}
\end{figure*}

Pre-SN radial density profiles of both 3D simulations along a single exemplary ray are shown in Fig. \ref{fig:pre_CSM}. Both simulations have the expected freely expanding and shocked wind phases, separated by a WTS. These profiles are generally representative of the features in both simulations, but can be more complex for other choices of direction, e.g. multiple shells will be present for the WR model along some directions. Due to the lower mechanical luminosity during the RSG phase, the density inside the RSG model's freely expanding wind is approximately two orders of magnitude greater than that of the WR model. The RSG wind's density initially jumps at the WTS by the expected factor of four, but increases further with distance downstream due to radiative cooling. The density increases by a factor $\approx 30$ beyond the WTS. In contrast to previous work (e.g. Fig. 2 of \citet{DWARKADAS2005} and \citet{Yasuda2021}), we do not produce a thin dense shell followed by a lower density MS bubble. Instead, we observe an extended region of increasing density over $\sim$$3\,$pc, outside of which is the ISM.
This is due to our assumption of a photoionized wind and not from omitting the MS wind, as we obtain similar results for our 2D CSM evolution including MS wind (see upper right panel of Fig. \ref{fig:2D_bubble}). When the wind and ISM are photoionized and heated to $\sim8000$\,K, the Mach number of the shocks induced by the RSG wind is much reduced, compression factor decreased, shell thickness increased and the outward-moving forward shock dissipates into a sound wave.
The RSG wind bubble remains close to spherical because its expansion is subsonic and the almost-isotropic external pressure (thermal pressure dominates over dynamic and magnetic) results in an almost spherical bubble.

In both simulations, the shocked wind gas displaces ISM and stellar wind from previous evolutionary phases. In the RSG simulation this is subsonic expansion into the ISM, and so the density changes only by $\sim$20\%. In the WR simulation, the faster wind of the WR star sweeps out the residual RSG gas into a dense shell (see Fig. \ref{fig:2D_bubble}), which expands supersonically at $\sim100$\kms. The dense shell is radiative and Rayleigh-Taylor (RT) unstable. The magnetic field strength inside the shell reaches peak values of 10s of $\mu$G. The dense shell driven by the WR wind is also aspherical because of the confining effect of ISM magnetic pressure.

The radial CSM profiles in our simulations differ from those presented in \cite{GarciaSegura1996a_RSG_WR} for both cases. In the RSG case, they produce a thin shell at the WTS at 3 pc (slow RSG wind) or 10 pc (fast RSG wind) pc, surrounded by a low-density bubble and another shell from their MS. 
We attribute the lack of RSG shell in our simulation to our assumption of external photoionization, which heats the RSG wind and the ISM, reducing the Mach number (and compression factor) of the WTS and rendering subsonic the expansion into the ISM.

Our RSG WTS is at a comparable radius to their slow wind case. Direct comparison with their WR simulation is difficult due to the lack of an equivalent radial density plot. However, comparing with their Fig. 7a, qualitatively there are some differences. In their work, they have a thin WR shell, which they artificially perturb with 1\% noise. They observe Vishniac instabilities \citep{Vishniac1983} as the dominant source of clumping, with some RT fingers as well. In our simulation, as we have wrapped a 2D CSM to 3D, we observe clumping and dynamical instabilities in the shell only in the XZ plane, as the wrapping is done in the XY plane.

The qualitative radial structure of our single star WR CSM is similar to that found by \citet{Yasuda2022}. They model a binary Ib/c CCSN progenitor, where dense and slow-moving Roche Lobe Overflow (RLOF) material is swept out by the subsequent WR wind, similar to the RSG material in our WR simulation. Their WR WTS and swept-up shell are both at approximately twice the radius compared to our simulation, and their swept-out shell appears to be thinner than ours. We attribute these differences to the differing stellar evolution assumptions and lack of hydrodynamic instabilities in their 1D treatment.

\subsection{RSG-SNR Simulation}
\label{subsec:RSG_results}
\subsubsection*{Hydrodynamic results}

\begin{figure}
    \centering
    \includegraphics[width=\linewidth]{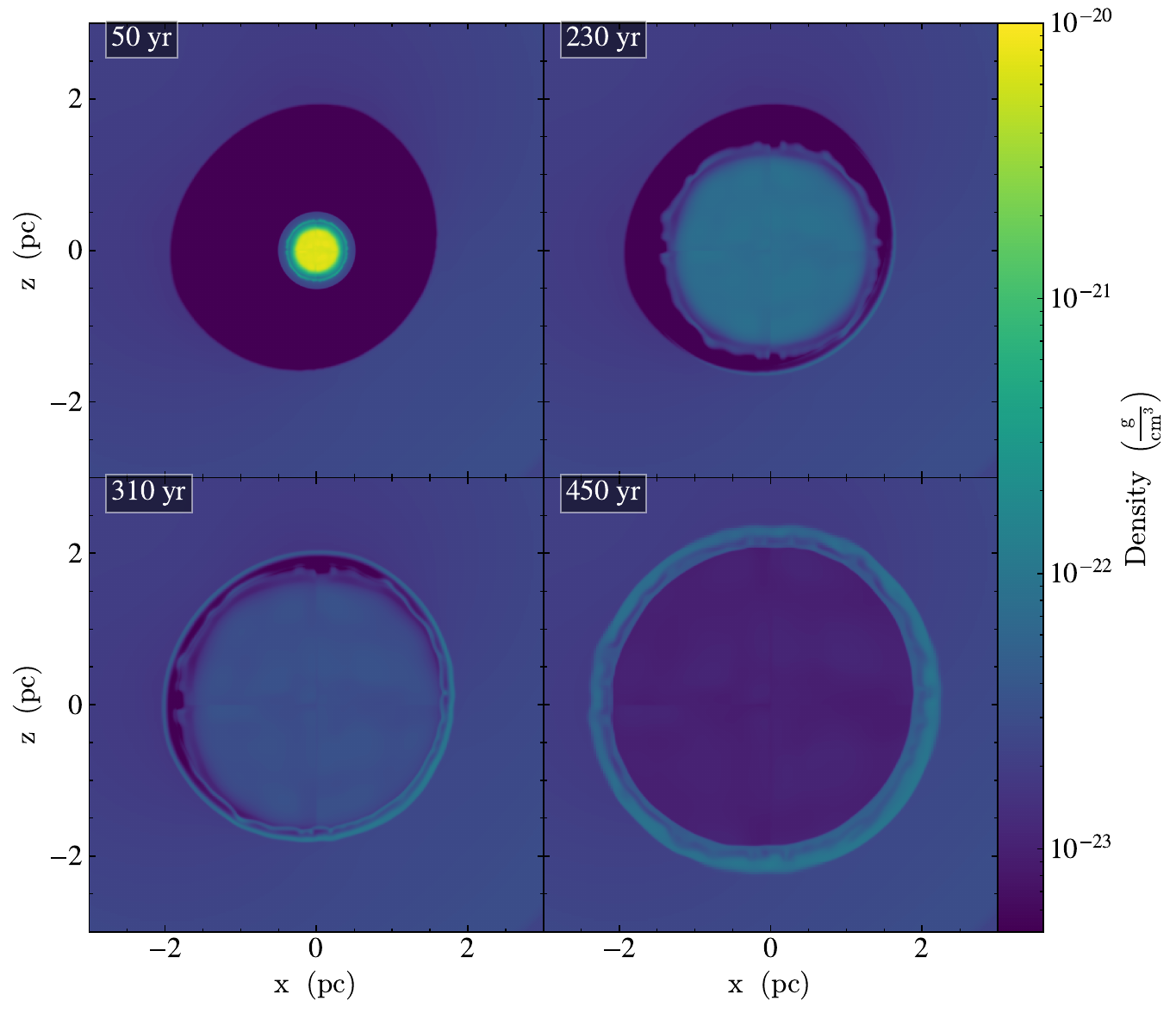}
    \caption{XZ-plane slice of density for selected times of the RSG simulation.}
    \label{fig:RSG_XZ_density}
\end{figure}

\begin{figure}
    \centering
    \includegraphics[width=\linewidth]{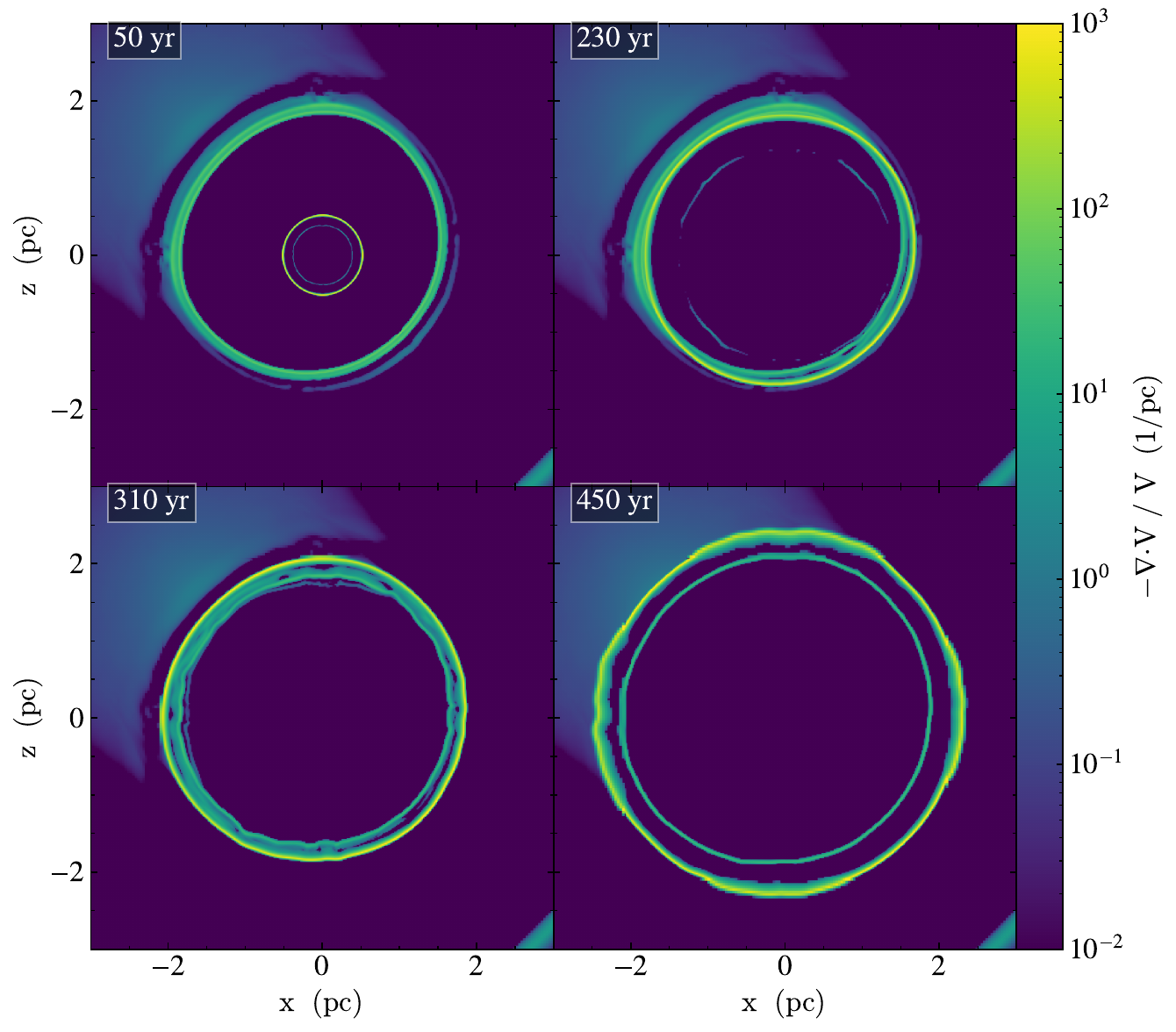}
    \caption{XZ-plane slice of $-\nabla\cdot \mathbf{V} / \mathrm{V}$ for selected times of the RSG simulation.}
    \label{fig:RSG_XZ_negdivvV}
\end{figure}

We evolve the simulation for 550\,yr post-explosion, by which time the SNR has expanded deep into the shocked RSG wind region. We can describe the evolution of the SNR in four phases, as shown in the four panels of Fig. \ref{fig:RSG_XZ_density} and Fig. \ref{fig:RSG_XZ_negdivvV} which are slices of density and $-\nabla\cdot \mathbf{V} / \mathrm{V}$ respectively. 

The SNR initially expands freely into the RSG stellar wind, with its $1/r^2$ density profile. The stellar WTS is compressed in the direction of proper motion. The initial aspherical profile imposed in the SN explosion is quickly smoothed out as the remnant expands. A strong forward shock is established at the interface with the CSM, and a reverse shock builds up over time ($\sim$$50$\,yr). The SNR sweeps up the RSG wind material with an approximately constant velocity, until it reaches the RSG WTS at about 1.7 pc ($\sim$$230$\,yr). 
At this point the SNR has swept up an amount of mass comparable to that of the ejecta $\sim$5\,\Msun\ and continues to expand. By t$~\sim$ 300\,yr, the WTS has been fully swept up and the forward and reverse shocks weaken and diverge. They form two clear shells corresponding to the two shocks ($\sim$$450$\,yr). At this point, the remnant is expanding into an approximately constant-density medium. Eventually, the reverse shock is expected to return to near the centre of explosion \citep[e.g.][]{DWARKADAS2007}. 

\subsubsection*{Magnetic field results}

\begin{figure}
    \centering
    \includegraphics[width=\linewidth]{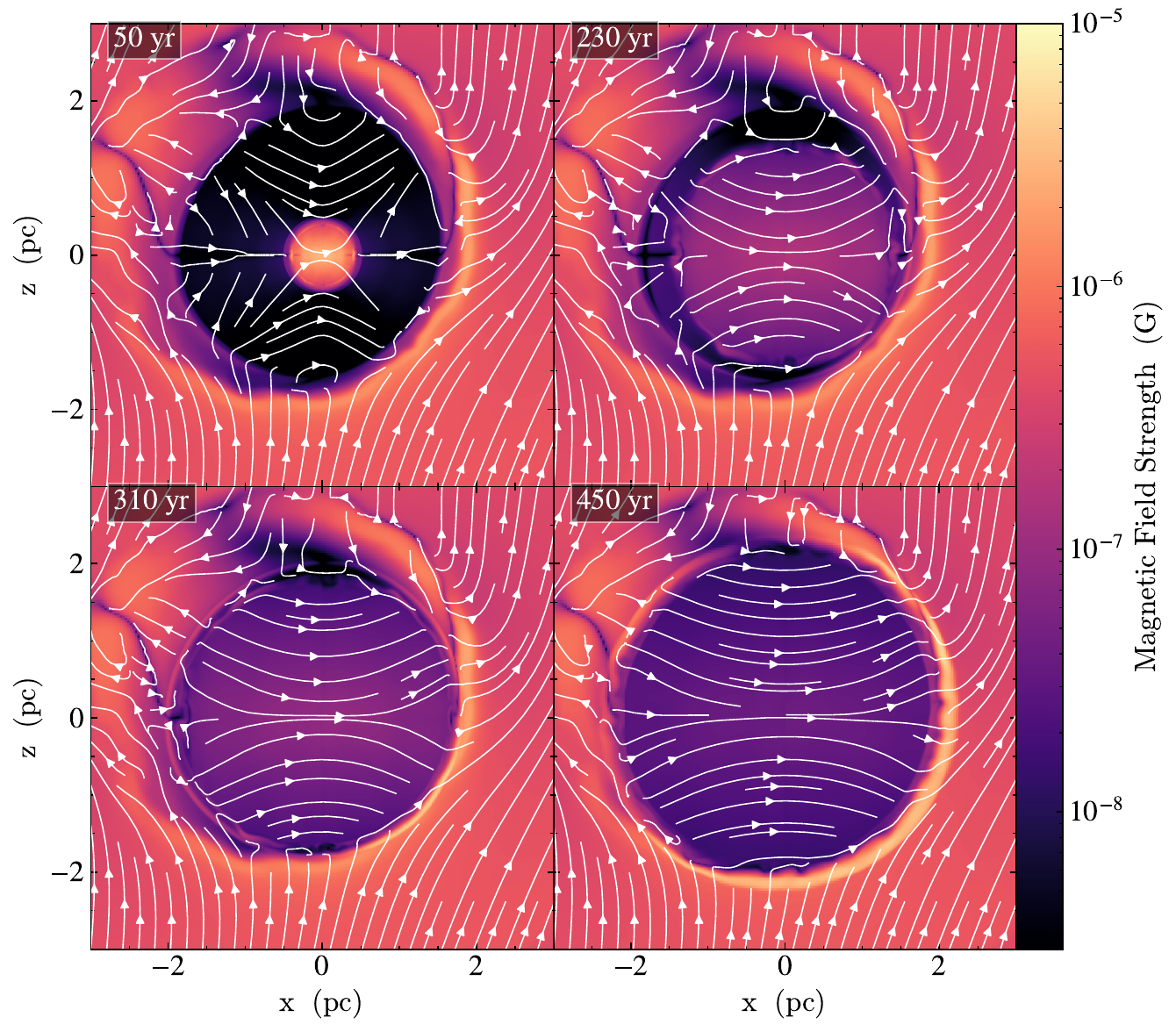}
    \caption{XZ-plane slice of magnetic field strength and streamlines of in-plane magnetic field direction for selected times of the RSG simulation.}
    \label{fig:RSG_XZ_B}
\end{figure}

In Fig.\,\ref{fig:RSG_XZ_B} we show the evolution of the magnetic field strength and orientation in a slice through the simulation. Due to the non-zero ISM velocity and slow RSG wind, a global asymmetry results leading to the SNR's forward shock reaching the WTS at different times. The magnetic field strength increases due to compression of the frozen-in field by cooling. As the WTS is swept up, the magnetic field is compressed, reaching peak values of order $10\,\mu$G, localised in a shell of $< 0.5$pc thickness.

\subsection{WR-SNR Simulation}
\label{subsec:WR_results}
\subsubsection*{Hydrodynamic results}
The WR scenario evolves the SNR for 2000\,yr post explosion. The longer simulation time is necessary to observe the effects of the WR CSM, which is a factor $\sim 2$ more spatially extended than the RSG case. 

\begin{figure}
    \centering
    \includegraphics[width=\linewidth]{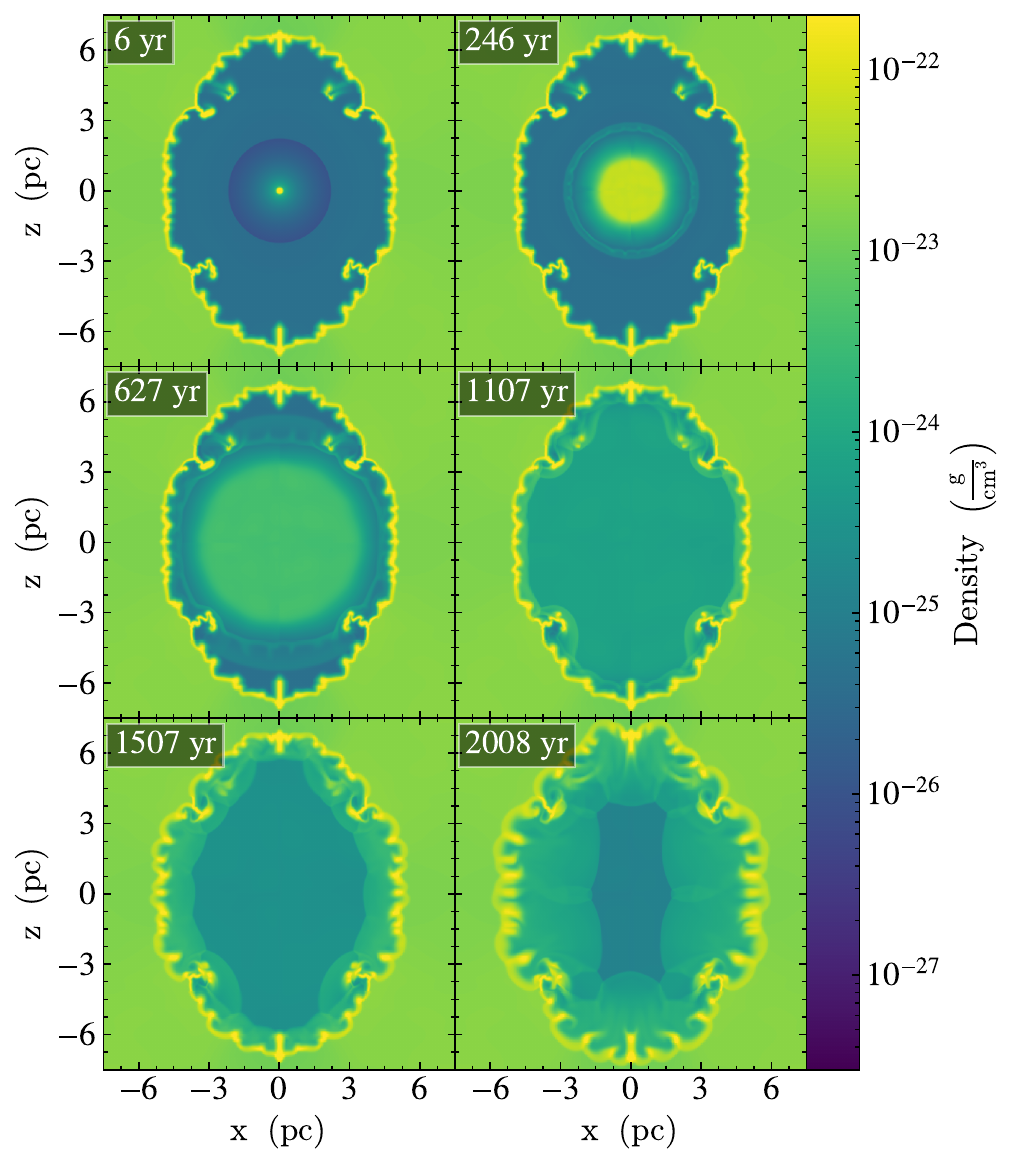}
    \caption{XZ-plane slice of density for selected times of the WR simulation.}
    \label{fig:WR_XZ_density}
\end{figure}

\begin{figure}
    \centering
    \includegraphics[width=\linewidth]{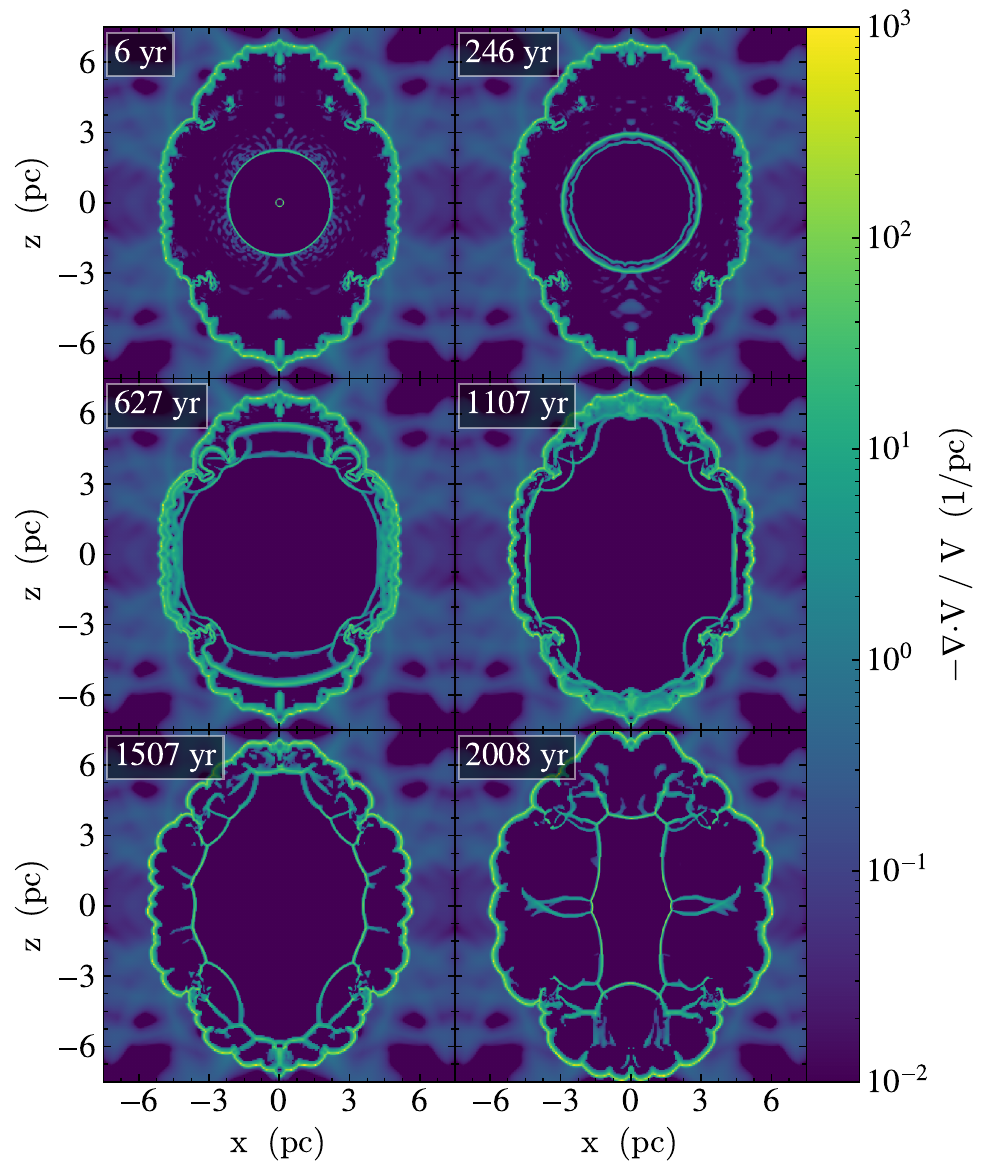}
    \caption{XZ-plane slice of $-\nabla\cdot \mathbf{V} / \mathbf{V}$ for selected times of the WR simulation.}
    \label{fig:WR_XZ_negdivvV}
\end{figure}

We describe the evolution of the SNR as shown in the six panels of Fig. \ref{fig:WR_XZ_density} and Fig. \ref{fig:WR_XZ_negdivvV} which are slices of density and $-\nabla\cdot \mathbf{V} / \mathrm{V}$ respectively.
The WR simulation begins similarly to the RSG simulation, with a free expansion phase for the first $\sim$ $200$\,\yr\, with the SN explosion asphericity smoothing out rapidly before reaching the WR WTS. Unlike in the RSG case, deceleration of the SNR shocks is minimal at this point due to the lower density in the freely expanding wind. After sweeping up the WR WTS, the SNR shocks continue to freely expand in the shocked wind, which is also less dense than in the RSG case. The SNR continues to expand until it hits the swept-up shell of dense RSG material at $\sim$ $600$\,\yr\,. The forward shock decelerates abruptly and reflected components travel backwards in the centre-of-explosion frame and interact with the reverse shock at $\sim$ 1000\,yr. The shock structure and dynamics at the RSG shell are complex until around 1400\,yr post-explosion, when a coherent set of reflected shocks detach from the shell and move inwards towards the centre of explosion. As they approach the origin, a contact discontinuity develops at the RSG shell, and the forward shock continues to expand. At 2000\,yr post-explosion, some of the reflected shocks are within 2 pc of the origin.

\subsubsection{Magnetic field results}

\begin{figure}
    \centering
    \includegraphics[width=\linewidth]{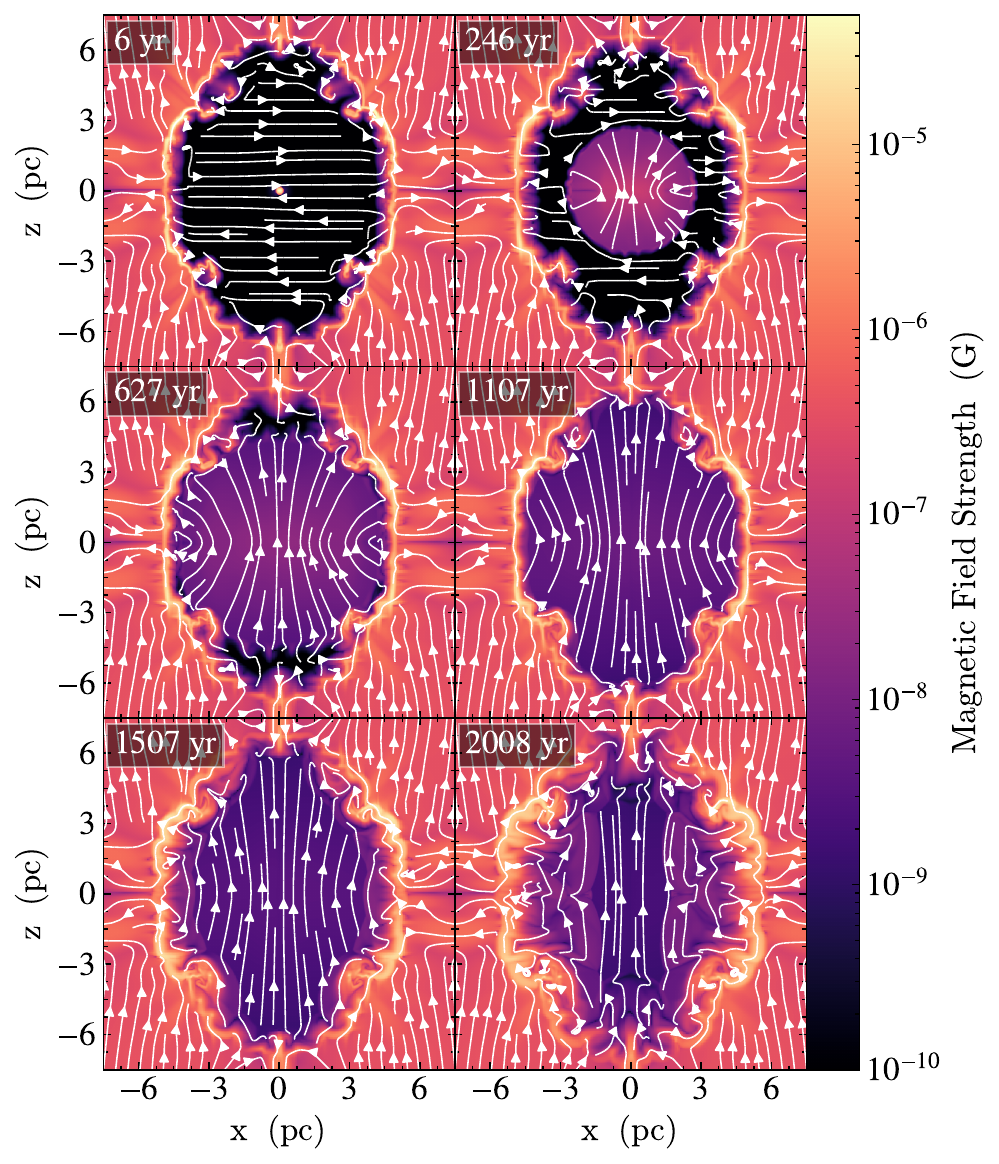}
    \caption{XZ-plane slice of magnetic field strength and streamlines of in-plane magnetic field direction for selected times of the WR simulation.}
    \label{fig:WR_XZ_B}
\end{figure}

In Fig. \ref{fig:WR_XZ_B} we show the evolution of the magnetic field strength and orientation in the XZ plane over time. 
In the WR simulation, the impact of including the RSG evolution can be clearly seen in the density and magnetic field structures. The SNR shock has a dense shell of RSG wind material to hit, favourable for accelerating particles, and the magnetic field strength reaches peak values of order a few $10^{-5}$ G due to radiative compression. The position of the RSG shell is relevant for potential particle acceleration, as the shock velocity decreases with shell radius (see Fig. \ref{fig:VshRsh_WR}). This radius will be determined by the stellar wind parameters and duration of the WR phase, as well as the external environmental conditions.

\subsection{Shock properties}
\label{subsec:shock_results}
In this section we examine the properties of the forward shocks in our simulations and compare with theoretical predictions. We adapted the SNR evolutionary model calculator of \cite{Leahy2017}, which uses the analytic solutions of \cite{TrueloveMcKee1999} for the timescales of interest here.

In each case we compare from the earliest timestep until the last timestep before the forward shock encounters the WTS, where the assumptions in \cite{TrueloveMcKee1999} are no longer valid. Their model requires a constant \mdot\ and \vinf\ to be assumed, so we make two comparisons. The first model takes the average of these parameters over the full duration of the preceding evolutionary stage (referred to as ``average'') in the \textsc{MESA} track. The second model takes the average of these parameters over only the duration of the advection time for the preceding evolutionary stage in the \textsc{MESA} track, this being the last $\sim 60$\,kyr for the RSG stage and the last $\sim 1$\,kyr for the WR stage (referred to as ``late''). Given the finite size $r_{max}$ of the SN when initialised in our simulations, we shift the model radius values by a constant such that at $t_{max}$ the radius is equal to $r_{max}$.

\begin{figure}
    \centering
    \includegraphics[width=1\linewidth]{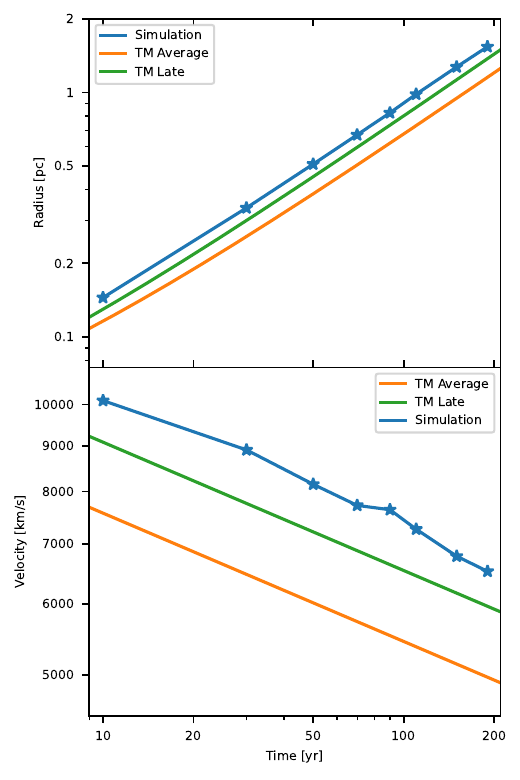}
    \caption{Evolution of RSG simulation forward shock radius (upper panel) and shock velocity (lower panel) vs time up to the WTS. The profile is calculated along an exemplary diagonal ray from the origin in the $[+x,-y,+z]$ direction. Predictions from the model of \cite{TrueloveMcKee1999} are also shown for the average and late cases as discussed in the main text. All velocities are in the centre-of-explosion frame.}
    \label{fig:VshRsh_RSG}
\end{figure}

In Fig. \ref{fig:VshRsh_RSG} we plot the position and velocity of the forward shock in our RSG simulation inside the WTS as a function of time, alongside the average and late \cite{TrueloveMcKee1999} model predictions for a steady state stellar wind. Within this freely expanding wind region, the radial density and velocity profile of the CSM is spherically symmetric for both simulations. We show profiles along a single exemplary ray, which is representative of all directions inside the WTS. Note that the shock velocity $V_{\mathrm{sh}} = \frac{dR_{\mathrm{sh}}}{dt}$ is measured in the lab frame. 

\begin{figure}
    \centering
    \includegraphics[width=1\linewidth]{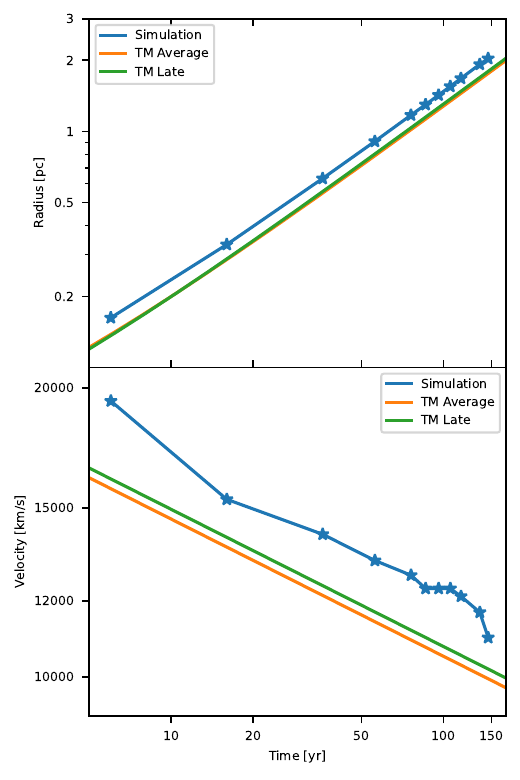}
    \caption{As Fig. \ref{fig:VshRsh_RSG} for the WR simulation.}
    \label{fig:VshRsh_WR}
\end{figure}

The average values of \mdot\ = $3\times 10^{-5}$ \Msun \pyr\ and \vinf\ = 28 \kms, which are typical values for RSGs, lead to underestimates of the shock position and velocity. The late values of \mdot = $1.5\times 10^{-5}$ \Msun \pyr and \vinf = 50 \kms are closer to the values in the simulation. In the RSG case, we observe some additional wind acceleration from thermal pressure due to assuming photoionization. In Fig. \ref{fig:VshRsh_WR} we do the same for the WR simulation.

Both sets of stellar wind values (average: \mdot = $2.8\times 10^{-5}$ \Msun \pyr and \vinf = 2540 \kms, late: \mdot = $2.8\times 10^{-5}$ \Msun \pyr and \vinf = 3100\kms) underestimate the radius and velocity, but only by 10-25\%. We note that \cite{TrueloveMcKee1999} assume the CSM itself is static, which our results suggest leads to underestimation of the resulting shock velocities. The wind parameters in the WR stage are also more variable than during the RSG stage, and so the assumptions of a steady state and static CSM by \cite{TrueloveMcKee1999} may not be as tenable for this simulation. 

\section{Discussion}
\label{sec:discussion}

\subsection{Stellar evolution}
\label{subsec:discussion_stevol}

In the context of particle acceleration calculations, the CSM is often approximated using an average or optimistic value for \mdot\, and \vinf. This approach is appropriate if only the freely-expanding wind phase is being considered. Beyond the freely-expanding wind regime, there is no such thing as a ``typical'' CSM in terms of number, position and magnitude of features in the CSM density profile \citep[see e.g. Fig. 4 of][]{FICHTNER2024}. Different stellar evolution scenarios can generate a wide variety of different CSM density profiles. These features will mostly depend on the initial mass and multiplicity of the progenitor, and their effects on SNe have been considered by e.g.\citep{BROSE2025,Das2024,MEYER2024}. The latest stages of massive star evolution are the least certain, but intense, asymmetric and episodic mass-loss is expected in many cases. Given the fast WR wind advection time, we would expect qualitative differences from different WR wind prescriptions to be small, unless \vinf\ decreases significantly over several tens of kyr and/or \mdot\ changes rapidly on similar timescales. Runaway \mdot\ at the end of the RSG phase \citep[e.g.][]{KEE2021,Bronner+2025} would likely lead to significant deviations from analytical predictions based on time-averaged steady-state values for \vinf\ and \mdot.    

It is not uncommon that values for \mdot, \vrot\ and $B_\star$ are assumed independently. However, high surface fields will impede mass-loss, and high mass-loss will lead to decreasing \vrot. Using a stellar evolution model as a basis for the CSM accounts for these values simultaneously, and thus provides a greater degree of self-consistency. The lack of rapid rotation implies a radially-dominated magnetic field that falls off rapidly as $B \propto 1/r^2$ in this case. In this work we assume a stellar evolution track with a modest initial \vrot\ consistent with observed Galactic O stars \citep{Holgado2022}. A fully non-rotating star would be an even less promising candidate for accelerating cosmic rays to PeV energies, which is why we did not model this scenario. Modifying the initial \vrot\ will also affect a star's evolutionary path, and thus the CSM and whether it explodes as a CCSN or not. Related to this, we only consider the ``classical'' WR evolutionary scenario with single star evolution. Other formation channels for stripped-envelope SN progenitors have been suggested, invoking effects such as mass-transfer and common-envelope evolution which would affect the CSM. Their impact on SNRs have been considered by e.g. \citep{Ercolino2024a,Ercolino2024b}. The structure of our WR CSM is qualitatively similar to that obtained by \citet{Yasuda2022} for a binary WR scenario, suggesting inclusion of shells of swept-up material from e.g. RSG and/or RLOF phases are important for hydrodynamic simulations of Type Ib/c in general. 

In this work we are testing stellar evolution predictions by calculating the CSM self-consistently up to the pre-SN stage, and then implementing simple explosion models to produce SNRs. This is qualitatively different to the work of \citet[e.g.,][]{Orlando2021,Orlando2025GM,Orlando20251987}, where detailed, tailored explosion and CSM prescriptions are used to test whether they reproduce observations of specific SNRs. 

\subsection{Stellar environment}
\label{subsec:discussion_environment}
Most massive stars are found in clusters. The external density and pressure from the environment can strongly affect the length scales of a star's CSM. For example, recent JWST images of the massive young cluster Westerlund 1 show RSG winds being confined by winds of other cluster members \citep{Guarcello2024}. RSGs in clusters are also likely to have photoionized winds for the same reason \citep{Mackey2015}, and we show here that this leads to thicker shells with smaller compression ratios, affecting the subsequent evolution of the SNR. RSGs embedded in a collective cluster wind can have their own wind ablated, leading to a highly asymmetric CSM \citep{Larkin25}. CCSNe in clusters will explode into a CSM with qualitatively different properties from those in the field, and are thought to be promising sites for production of multi-PeV protons \citep{Vieu22}. 

A minority of massive stars are isolated from clusters. While it is debated whether these stars are forming in-situ \citep[e.g.,][]{Oey2013,Oskinova2013} or are runaways from clusters \citep[e.g.,][]{Gvaramadze2012,Vargas-Salazar2020}, in either case the CSM of these stars will be qualitatively different due to the lower external densities and pressures they encounter in the field vs in a cluster. Already in our RSG simulation, the impact of a small 3D peculiar velocity is apparent in the density and magnetic field structure both pre- and post-SN. We see a compression in the WTS in the direction of motion, causing the SN to reach the WTS in this direction sooner. In the case of runaway stars with high peculiar velocities, their wind-ISM interaction produces a bow shock as the star sweeps up ISM material. This produces a strongly asymmetric CSM, as considered for example by \citet{Meyer2015,Meyer2017}.

\subsection{SNR evolution}
\label{subsec:discussion_post_SN_evol}
Analytic prescriptions no longer hold in the region beyond the WTS, and the effects of non-linear density structures beyond this point need to be probed via simulations. In particular, the reflected shock in our WR model processes a significant amount of energy ($\sim3\times10^{50}$ erg), but only occurs due to the presence of RSG shell material that is often not included in simulations \citep[e.g.,][]{Telezhinsky2013,ZP2018}. Even within the WTS, we find shock velocities faster than predicted by \citet{TrueloveMcKee1999} in both cases. For the RSG case, photoionization produces an additional wind acceleration, and for the WR case the expansion velocity of the CSM itself is high enough to become relevant.

A consequence of our 3D treatment is our relatively poor spatial resolution at pc scales. At certain points we cannot fully resolve the complex set of interactions between different reflected and transmitted shocks. However it is clear that the forward shocks are still moving with velocities in excess of 1000\,\kms\, after encountering the WTS in both cases, and therefore may be sites of non-thermal emission.

In both of our simulations, we observe enhanced magnetic field compression in the CSM from cooling. The Axford-Cranfill effect may be occuring, but given the field only increases by a factor $\sim2-3$ it is not noticeable compared with the cooling-induced compression. The finding of $B\sim r$ within the stellar bubble in \cite{ZP2018} is not reproduced here. The effect is also asymmetric, particularly in our RSG simulation due to the small proper motion.

Our inclusion of an aspherical explosion to break artificial symmetries did not appear to significantly affect the SNR evolution in either case, as these features were found to smooth out rapidly. A more extreme asymmetry such as a bipolar eruption would produce significant deviations, as found by e.g. \citet{Orlando20251987}.

\subsection{Implications for particle acceleration}
\label{subsec:discussion_PA}
The output of the simulations allow us to consider the implications for particle acceleration, and associated non-thermal emission in young SNRs. To find the shock velocity $\varv_\text{sh}$ in the simulations\footnote{Note that $\varv_\text{sh}$ is the flow velocity through the shock in the reference frame of the shock, as opposed to $V_{\mathrm{sh}}$ defined above, which is how fast the shock is expanding in the lab frame.}, we use a shock locating algorithm to determine the shock locations, and calculate the relative velocity between the fluid immediately upstream and downstream of the shock surface. The shock velocity is then
\begin{equation}
    \varv_\text{sh} = \frac{r}{r-1} \left( \varv_{\mathrm{DS}} - \varv_{\mathrm{US}} \right)\text{,}
\end{equation}
where $r$ is the compression ratio $\rho_{\mathrm{DS}}/\rho_{\mathrm{US}}$, and $\varv_{\mathrm{DS}}$ and $\varv_{\mathrm{US}}$ are the radial velocities downstream and upstream of the shock respectively. Defining $\varv_\text{sh}$ in this way is convenient when calculating the energy processed by the shock per unit time per unit area is
$\frac{dE}{dt dA} \approx \frac{1}{2} \rho_\mathrm{US} \varv_\text{sh}^3, $
where $dA$ is a differential area element of the shock's surface. This quantity is useful in determining the likelihood of cosmic-ray driven magnetic field amplification at the different shocks.

\subsubsection*{Forward shocks}
For both the WR and RSG simulations, the shock initially expands into the CSM of the progenitor. As discussed it is the late time stellar evolution, i.e. in the ``immediate'' pre SN period, that sets the shock and wind conditions. In this case, we can apply the method of \cite{BELL2013} \cite[see also][]{2022Sci...376...77H}, to determine the maximum energy in the early phase of evolution.

We assume some fraction $\eta_{\rm esc}$ of the energy processed by the shock is converted to energetic protons (cosmic rays) which escape upstream, driving the growth of self-confining magnetic fluctuations as they do so \cite[see][for details]{BELL2013}. For the forward shock, propagating in a $1/r^2$ wind profile, the predicted maximum particle energy is 
\begin{align*}
  E_{\rm max}^{\rm (FS)} \approx \left(\frac{\eta_{\rm esc}}{10^{-2}}\right)
   \left(\frac{\dot{M}}{10^{-5}\,{\rm M}_\odot~{\rm yr}^{-1}}\right)^{\frac{1}{2}}
    \left(\frac{\varv_{\infty}}{10{\kms}}\right)^{-{\frac{1}{2}}}
     \left(\frac{\varv_{\rm sh}}{10^3{\kms}}\right) ^2
     {\rm TeV}
\end{align*}

The above estimate assumes the escaping current produced by the accelerated protons is sufficient to drive the non-resonant current driven instability described in \citet{Bell04}. For the simulation results presented here, i.e. using the late wind parameter values as described earlier, this equates to maximum energies in the first decade post SN of $\approx 100$\,TeV and $\approx 30$ TeV for the RSG and WR cases respectively. Due to the evident shock deceleration the maximum energy reduces with time. 

The external shocks will eventually reach the zone where enhanced magnetic field compression occurs, either due to cooling in the immediate post-shock region, or more gradually, via the Axford-Cranfill effect, should it operate. In contrast to \citet{ZP2018}, we find the enhanced field regions are localised to a narrow layer beyond the transition where cooling drives strong compression. However, given the modest velocities $\approx 1000$ \kms\ the shock retains at these locations, any effect on the maximum energy is minor. This may be due to the fact that our time-evolving value for \vrot\ is approximately an order of magnitude lower than that of \citet{ZP2018} immediately pre-SN, and thus $r_t$ is further from the star (see panel (c) of Fig. \ref{fig:Bevol}). It is evident that extreme wind conditions, e.g.\ high mass loss rates with slow wind, and/or SNR conditions e.g. extreme shock velocities, are required for the acceleration of protons above PeV energies at the external shocks in isolated SNRs.

\subsubsection*{Reflected shocks}

Since the reflected shock in the WR case processes a substantial amount of energy, we should consider here the possible radiative signatures of such a shock. Owing to the large expansion of the gas, the magnetic field strength internal to the reflected shock in the simulations is weak, $B\ll \mu$G. A minimum requirement for excitation of the non-resonant instability is that $B<B_\text{NR}$ \citep{Bell04}, where 
\begin{align*}
 B_\text{NR} =  \left(\frac{f_{\rm esc}}{10^{-2}}\right)^{\frac{1}{2}}
    \left(\frac{\rho}{10^{-25} {\rm g~cm}^{-3}}\right)^{\frac{1}{2}}
     \left(\frac{\varv_{\rm sh}}{5000\,{\kms}}\right) ^{\frac{3}{2}} \,{\rm \mu G}.  
\end{align*}
Here $f_{\rm esc}$ is the fraction of energy processed by the shock escaping upstream as cosmic rays flow into the SNR interior. The numerical values are motivated from the simulation results, and evidently, $B_{\rm sim} \ll B_\text{NR}$, implying the non-resonant instability is important for magnetic field amplification at the reflected shock.

In the case of a reflected shock, its surface is converging as it moves inward, the escaping particles will thus focus towards the origin. While this may play an interesting role, a detailed model of this is beyond the scope of this work. To keep things simple, we adopt the most optimistic scenario, where the field is  amplified to $B_\text{NR}$, and assume particles reach the Hillas limit in this field. We thus set an upper limit to the maximum particle energy
\begin{align*}
    E_{\rm Hillas} &= q B R u_{\rm sh}/c \\
    &\approx 20Z\left(\frac{f_{\rm esc}}{10^{-2}}\right)^{\frac{1}{2}}
    \left(\frac{\rho}{10^{-25} {\rm g~cm}^{-3}}\right)^{\frac{1}{2}}
     \left(\frac{\varv_{\rm sh}}{5000\,{\kms}}\right) ^{\frac{5}{2}} \left(\frac{R_{\rm sh}}{1\,{\rm pc}}\right) \,{\rm TeV}
\end{align*}
where again, numerical values are motivated from the simulation.

The implied low magnetic field values and modest maximum energies essentially rule out the possibility of non-thermal X-ray emission.
Since the cooling time for electrons is
\begin{align*}
    t_{\rm cool} \approx 300 \left(\frac{E}{1\,{\rm TeV}}\right)^{-1} 
    \left(\frac{u_{\rm ph} + u_{B}}{1\,{\rm eV\, cm}^{-3}}\right)^{-1} 
    \, {\rm kyr}
\end{align*}
any population accelerated by the reflected shock will live long after the reflected shock has dissipated. 

Let us assume the electrons are accelerated into a power-law energy distribution with index $s \gtrsim 2$, and a fraction $\eta_e\approx 0.01\%$ of the total energy processed is given to electrons above GeV energies. In the simulation, the reflected shock processes approximately $3\times 10^{50}$\,erg. If these particles are confined in the SNR interior, they may produce detectable emission in gamma rays. To estimate the gamma-ray flux, we approximate the target radiation field as a monochromatic population of photons with energy 0.1 eV with uniform energy density of 1 eV / cm$^{3}$. The resulting inverse-Compton luminosity for times $t_{\rm SNR} \ll t_{\rm cool}$ is $L(E>{\rm TeV}) \approx 10^{32} ~{\rm erg\, s}^{-1}$, which would be detectable with current generation imaging atmospheric Cherenkov telescopes if the source was within a few kpc distance. 

\subsubsection*{Prospects for CCSNe as PeVatrons}
We find maximum particle energies of order a few tens of TeV for all cases considered, notably lower than was found by \citet{ZP2018} for the case of a WR progenitor for the reasons discussed above.
This is broadly in line with other recent works \citep[e.g.][]{BROSE2022}, which find that extreme conditions are required to achieve PeV energies. Such possibilities include a powerful SN inside a star cluster \citep[e.g.][]{Haerer2025}, successive SNe in a superbubble \citep[e.g.][]{Vieu22} or the SNR interacting with dense circumstellar material ejected shortly before explosion \citep[e.g.][]{BELL2013,BROSE2025}. Considering the reflected shock of the SNR in the WR progenitor case, the maximum energy of electrons is again found to be tens of TeV, with detectable inverse-Compton emission predicted.

\section{Conclusion}
\label{sec:conclusion}
In this work we computed 3D MHD simulations of young SNRs expanding through CSM generated self-consistently using a detailed stellar evolution treatment. We summarise our findings as follows:
\begin{enumerate}
    \item A multi-D treatment of the CSM is required to account for the effects of dynamical instabilities on the CSM structure.
    \item Our simulations suggest that analytic prescriptions assuming a steady-state and static CSM can underestimate forward shock velocities in young SNRs. For the RSG case, we find that photoionization introduces an additional acceleration to the stellar wind, and for the WR case the CSM expansion velocity is also important.
    \item Our simulations also suggest that using analytic predictions calculated with stellar wind parameters time-averaged over a full evolutionary phase are likely to be discrepant with detailed simulations if the wind parameters change significantly towards the end of the star's life. We find time-averaging over the wind advection time to be a closer approximation in both of our cases.

    \item Detailed stellar evolution treatment produces CSM features beyond the free-expanding wind region, which the SNR will interact with as it expands.
    \item We show for a RSG that assuming full photoionization, as would be expected in a young stellar cluster, produces a qualitatively different CSM. We obtain an extended region of increasing density instead of a thin dense shell at the WTS.
    \item In agreement with previous works, we show that these interactions can produce favourable conditions for accelerating particles to a few tens of TeV, but are unlikely to reach PeV energies. We consider the particular case of a reflected shock propagating through a WR wind bubble for the first time, finding that inverse Compton emission may be detectable in gamma rays.
    \item We demonstrate the impact of using self-consistent stellar evolution models which ensure the \vrot\ value at explosion is consistent with the progenitor's mass-loss history.
    \item Cooling leads to magnetic field strengths of tens of $\mu$G at compressed regions, but we show this occurs only in thin shells, and the effects can be highly asymmetrical. We do not find the Axford-Cranfill effect to be contributing significantly in our simulations. 
    
\end{enumerate}

\begin{acknowledgements}

     We thank the referee for their constructive feedback which has improved this work. The simulations presented here were performed on the HPC system Raven at the Max Planck Computing and Data Facility. CJKL gratefully acknowledges support from the International Max Planck Research School for Astronomy and Cosmic Physics at the University of Heidelberg in the form of an IMPRS PhD fellowship.
     This publication results from research conducted with the financial support of Taighde \'Eireann - Research Ireland under Grant number 20/RS-URF-R/3712. 
     AACS is supported by the German \textit{Deut\-sche For\-schungs\-ge\-mein\-schaft, DFG\/} in the form of an Emmy Noether Research Group -- Project-ID 445674056 (SA4064/1-1, PI Sander). This project was co-funded by the European Union (Project 101183150 - OCEANS).
     This research made use of the following software packages: 
     Astropy \citep{astropy:2018},  Numpy \citep{HarMilVan20}, matplotlib \citep{Hun07}, yt \citep{TurSmiOis11}, \textsc{pion} \citep{2021PION}, \textsc{pypion} \citep{GreMac21}, \textsc{SNR.py} \citep{Leahy2017}.
\end{acknowledgements}

% WARNING
%-------------------------------------------------------------------
% Please note that we have included the references to the file aa.dem in
% order to compile it, but we ask you to:
%
% - use BibTeX with the regular commands:
   \bibliographystyle{aa} % style aa.bst
   \bibliography{biblio.bib} % your references 

@article{MESA_I,
       author = {{Paxton}, Bill and {Bildsten}, Lars and {Dotter}, Aaron and {Herwig}, Falk and {Lesaffre}, Pierre and {Timmes}, Frank},
        title = "{Modules for Experiments in Stellar Astrophysics (MESA)}",
      journal = {\apjs},
         year = 2011,
        month = jan,
       volume = {192},
       number = {1},
          eid = {3},
        pages = {3},
          doi = {10.1088/0067-0049/192/1/3},
archivePrefix = {arXiv},
       eprint = {1009.1622},
 primaryClass = {astro-ph.SR},
       adsurl = {https://ui.adsabs.harvard.edu/abs/2011ApJS..192....3P}
}

@article{MESA_II,
       author = {{Paxton}, Bill and {Cantiello}, Matteo and {Arras}, Phil and {Bildsten}, Lars and {Brown}, Edward F. and {Dotter}, Aaron and {Mankovich}, Christopher and {Montgomery}, M.~H. and {Stello}, Dennis and {Timmes}, F.~X. and {Townsend}, Richard},
        title = "{Modules for Experiments in Stellar Astrophysics (MESA): Planets, Oscillations, Rotation, and Massive Stars}",
      journal = {\apjs},
         year = 2013,
        month = sep,
       volume = {208},
       number = {1},
          eid = {4},
        pages = {4},
          doi = {10.1088/0067-0049/208/1/4},
archivePrefix = {arXiv},
       eprint = {1301.0319},
 primaryClass = {astro-ph.SR},
       adsurl = {https://ui.adsabs.harvard.edu/abs/2013ApJS..208....4P}
}

@article{MESA_III,
       author = {{Paxton}, Bill and {Marchant}, Pablo and {Schwab}, Josiah and {Bauer}, Evan B. and {Bildsten}, Lars and {Cantiello}, Matteo and {Dessart}, Luc and {Farmer}, R. and {Hu}, H. and {Langer}, N. and {Townsend}, R.~H.~D. and {Townsley}, Dean M. and {Timmes}, F.~X.},
        title = "{Modules for Experiments in Stellar Astrophysics (MESA): Binaries, Pulsations, and Explosions}",
      journal = {\apjs},
         year = 2015,
        month = sep,
       volume = {220},
       number = {1},
          eid = {15},
        pages = {15},
          doi = {10.1088/0067-0049/220/1/15},
archivePrefix = {arXiv},
       eprint = {1506.03146},
 primaryClass = {astro-ph.SR},
       adsurl = {https://ui.adsabs.harvard.edu/abs/2015ApJS..220...15P}
}

@ARTICLE{MESA_IV,
       author = {{Paxton}, Bill and {Schwab}, Josiah and {Bauer}, Evan B. and {Bildsten}, Lars and {Blinnikov}, Sergei and {Duffell}, Paul and {Farmer}, R. and {Goldberg}, Jared A. and {Marchant}, Pablo and {Sorokina}, Elena and {Thoul}, Anne and {Townsend}, Richard H.~D. and {Timmes}, F.~X.},
        title = "{Modules for Experiments in Stellar Astrophysics (MESA): Convective Boundaries, Element Diffusion, and Massive Star Explosions}",
      journal = {\apjs},
         year = 2018,
        month = feb,
       volume = {234},
       number = {2},
          eid = {34},
        pages = {34},
          doi = {10.3847/1538-4365/aaa5a8},
archivePrefix = {arXiv},
       eprint = {1710.08424},
 primaryClass = {astro-ph.SR},
       adsurl = {https://ui.adsabs.harvard.edu/abs/2018ApJS..234...34P}
}

@article{MESA_V,
       author = {{Paxton}, Bill and {Smolec}, R. and {Schwab}, Josiah and {Gautschy}, A. and {Bildsten}, Lars and {Cantiello}, Matteo and {Dotter}, Aaron and {Farmer}, R. and {Goldberg}, Jared A. and {Jermyn}, Adam S. and {Kanbur}, S.~M. and {Marchant}, Pablo and {Thoul}, Anne and {Townsend}, Richard H.~D. and {Wolf}, William M. and {Zhang}, Michael and {Timmes}, F.~X.},
        title = "{Modules for Experiments in Stellar Astrophysics (MESA): Pulsating Variable Stars, Rotation, Convective Boundaries, and Energy Conservation}",
      journal = {\apjs},
         year = 2019,
        month = jul,
       volume = {243},
       number = {1},
          eid = {10},
        pages = {10},
          doi = {10.3847/1538-4365/ab2241},
archivePrefix = {arXiv},
       eprint = {1903.01426},
 primaryClass = {astro-ph.SR},
       adsurl = {https://ui.adsabs.harvard.edu/abs/2019ApJS..243...10P}
}

@ARTICLE{2021PION,
	author = {{Mackey}, Jonathan and {Green}, Samuel and {Moutzouri}, Maria and {Haworth}, Thomas J. and {Kavanagh}, Robert D. and {Zargaryan}, Davit and {Celeste}, Maggie},
	title = "{PION: simulating bow shocks and circumstellar nebulae}",
	journal = {\mnras},
	year = 2021,
	month = jun,
	volume = {504},
	number = {1},
	pages = {983-1008},
	doi = {10.1093/mnras/stab781},
	archivePrefix = {arXiv},
	eprint = {2103.07555},
	primaryClass = {astro-ph.GA},
	adsurl = {https://ui.adsabs.harvard.edu/abs/2021MNRAS.504..983M}
}

@ARTICLE{BELL2013,
       author = {{Bell}, A.~R. and {Schure}, K.~M. and {Reville}, B. and {Giacinti}, G.},
        title = "{Cosmic-ray acceleration and escape from supernova remnants}",
      journal = {\mnras},
         year = 2013,
        month = may,
       volume = {431},
       number = {1},
        pages = {415-429},
          doi = {10.1093/mnras/stt179},
archivePrefix = {arXiv},
       eprint = {1301.7264},
 primaryClass = {astro-ph.HE},
       adsurl = {https://ui.adsabs.harvard.edu/abs/2013MNRAS.431..415B}
}

@ARTICLE{Vieu22,
       author = {{Vieu}, T. and {Reville}, B. and {Aharonian}, F.},
        title = "{Can superbubbles accelerate ultrahigh energy protons?}",
      journal = {\mnras},
         year = 2022,
        month = sep,
       volume = {515},
       number = {2},
        pages = {2256-2265},
          doi = {10.1093/mnras/stac1901},
archivePrefix = {arXiv},
       eprint = {2207.01432},
 primaryClass = {astro-ph.HE},
       adsurl = {https://ui.adsabs.harvard.edu/abs/2022MNRAS.515.2256V}
}

@ARTICLE{KEE2021,
       author = {{Kee}, N.~D. and {Sundqvist}, J.~O. and {Decin}, L. and {de Koter}, A. and {Sana}, H.},
        title = "{Analytic, dust-independent mass-loss rates for red supergiant winds initiated by turbulent pressure}",
      journal = {\aap},
         year = 2021,
        month = feb,
       volume = {646},
          eid = {A180},
        pages = {A180},
          doi = {10.1051/0004-6361/202039224},
archivePrefix = {arXiv},
       eprint = {2101.03070},
 primaryClass = {astro-ph.SR},
       adsurl = {https://ui.adsabs.harvard.edu/abs/2021A&A...646A.180K}
}

@ARTICLE{BROSE2022,
       author = {{Brose}, R. and {Sushch}, I. and {Mackey}, J.},
        title = "{Core-collapse supernovae in dense environments - particle acceleration and non-thermal emission}",
      journal = {\mnras},
         year = 2022,
        month = oct,
       volume = {516},
       number = {1},
        pages = {492-505},
          doi = {10.1093/mnras/stac2234},
archivePrefix = {arXiv},
       eprint = {2208.04185},
 primaryClass = {astro-ph.HE},
       adsurl = {https://ui.adsabs.harvard.edu/abs/2022MNRAS.516..492B}
}

@ARTICLE{BROSE2025,
       author = {{Brose}, R. and {Sushch}, I. and {Mackey}, J.},
        title = "{How to turn a supernova into a PeVatron}",
      journal = {\aap},
         year = 2025,
        month = jul,
       volume = {699},
          eid = {A160},
        pages = {A160},
          doi = {10.1051/0004-6361/202453334},
archivePrefix = {arXiv},
       eprint = {2504.20601},
 primaryClass = {astro-ph.HE},
       adsurl = {https://ui.adsabs.harvard.edu/abs/2025A&A...699A.160B}
}

@ARTICLE{ZP2018,
       author = {{Zirakashvili}, V.~N. and {Ptuskin}, V.~S.},
        title = "{Cosmic ray acceleration in magnetic circumstellar bubbles}",
      journal = {Astroparticle Physics},
         year = 2018,
        month = mar,
       volume = {98},
        pages = {21-27},
          doi = {10.1016/j.astropartphys.2018.01.005},
archivePrefix = {arXiv},
       eprint = {1712.02174},
 primaryClass = {astro-ph.HE},
       adsurl = {https://ui.adsabs.harvard.edu/abs/2018APh....98...21Z}
}

@ARTICLE{WHALEN2008,
       author = {{Whalen}, Daniel and {van Veelen}, Bob and {O'Shea}, Brian W. and {Norman}, Michael L.},
        title = "{The Destruction of Cosmological Minihalos by Primordial Supernovae}",
      journal = {\apj},
         year = 2008,
        month = jul,
       volume = {682},
       number = {1},
        pages = {49-67},
          doi = {10.1086/589643},
archivePrefix = {arXiv},
       eprint = {0801.3698},
 primaryClass = {astro-ph},
       adsurl = {https://ui.adsabs.harvard.edu/abs/2008ApJ...682...49W}
}

@ARTICLE{FermiSNRs,
       author = {{Ackermann}, M. and {Ajello}, M. and {Allafort}, A. and {Baldini}, L. and {Ballet}, J. and {Barbiellini}, G. and {Baring}, M.~G. and {Bastieri}, D. and {Bechtol}, K. and {Bellazzini}, R. and {Blandford}, R.~D. and {Bloom}, E.~D. and {Bonamente}, E. and {Borgland}, A.~W. and {Bottacini}, E. and {Brandt}, T.~J. and {Bregeon}, J. and {Brigida}, M. and {Bruel}, P. and {Buehler}, R. and {Busetto}, G. and {Buson}, S. and {Caliandro}, G.~A. and {Cameron}, R.~A. and {Caraveo}, P.~A. and {Casandjian}, J.~M. and {Cecchi}, C. and {{\c{C}}elik}, {\"O}. and {Charles}, E. and {Chaty}, S. and {Chaves}, R.~C.~G. and {Chekhtman}, A. and {Cheung}, C.~C. and {Chiang}, J. and {Chiaro}, G. and {Cillis}, A.~N. and {Ciprini}, S. and {Claus}, R. and {Cohen-Tanugi}, J. and {Cominsky}, L.~R. and {Conrad}, J. and {Corbel}, S. and {Cutini}, S. and {D'Ammando}, F. and {de Angelis}, A. and {de Palma}, F. and {Dermer}, C.~D. and {do Couto e Silva}, E. and {Drell}, P.~S. and {Drlica-Wagner}, A. and {Falletti}, L. and {Favuzzi}, C. and {Ferrara}, E.~C. and {Franckowiak}, A. and {Fukazawa}, Y. and {Funk}, S. and {Fusco}, P. and {Gargano}, F. and {Germani}, S. and {Giglietto}, N. and {Giommi}, P. and {Giordano}, F. and {Giroletti}, M. and {Glanzman}, T. and {Godfrey}, G. and {Grenier}, I.~A. and {Grondin}, M. -H. and {Grove}, J.~E. and {Guiriec}, S. and {Hadasch}, D. and {Hanabata}, Y. and {Harding}, A.~K. and {Hayashida}, M. and {Hayashi}, K. and {Hays}, E. and {Hewitt}, J.~W. and {Hill}, A.~B. and {Hughes}, R.~E. and {Jackson}, M.~S. and {Jogler}, T. and {J{\'o}hannesson}, G. and {Johnson}, A.~S. and {Kamae}, T. and {Kataoka}, J. and {Katsuta}, J. and {Kn{\"o}dlseder}, J. and {Kuss}, M. and {Lande}, J. and {Larsson}, S. and {Latronico}, L. and {Lemoine-Goumard}, M. and {Longo}, F. and {Loparco}, F. and {Lovellette}, M.~N. and {Lubrano}, P. and {Madejski}, G.~M. and {Massaro}, F. and {Mayer}, M. and {Mazziotta}, M.~N. and {McEnery}, J.~E. and {Mehault}, J. and {Michelson}, P.~F. and {Mignani}, R.~P. and {Mitthumsiri}, W. and {Mizuno}, T. and {Moiseev}, A.~A. and {Monzani}, M.~E. and {Morselli}, A. and {Moskalenko}, I.~V. and {Murgia}, S. and {Nakamori}, T. and {Nemmen}, R. and {Nuss}, E. and {Ohno}, M. and {Ohsugi}, T. and {Omodei}, N. and {Orienti}, M. and {Orlando}, E. and {Ormes}, J.~F. and {Paneque}, D. and {Perkins}, J.~S. and {Pesce-Rollins}, M. and {Piron}, F. and {Pivato}, G. and {Rain{\`o}}, S. and {Rando}, R. and {Razzano}, M. and {Razzaque}, S. and {Reimer}, A. and {Reimer}, O. and {Ritz}, S. and {Romoli}, C. and {S{\'a}nchez-Conde}, M. and {Schulz}, A. and {Sgr{\`o}}, C. and {Simeon}, P.~E. and {Siskind}, E.~J. and {Smith}, D.~A. and {Spandre}, G. and {Spinelli}, P. and {Stecker}, F.~W. and {Strong}, A.~W. and {Suson}, D.~J. and {Tajima}, H. and {Takahashi}, H. and {Takahashi}, T. and {Tanaka}, T. and {Thayer}, J.~G. and {Thayer}, J.~B. and {Thompson}, D.~J. and {Thorsett}, S.~E. and {Tibaldo}, L. and {Tibolla}, O. and {Tinivella}, M. and {Troja}, E. and {Uchiyama}, Y. and {Usher}, T.~L. and {Vandenbroucke}, J. and {Vasileiou}, V. and {Vianello}, G. and {Vitale}, V. and {Waite}, A.~P. and {Werner}, M. and {Winer}, B.~L. and {Wood}, K.~S. and {Wood}, M. and {Yamazaki}, R. and {Yang}, Z. and {Zimmer}, S.},
        title = "{Detection of the Characteristic Pion-Decay Signature in Supernova Remnants}",
      journal = {Science},
         year = 2013,
        month = feb,
       volume = {339},
       number = {6121},
        pages = {807-811},
          doi = {10.1126/science.1231160},
archivePrefix = {arXiv},
       eprint = {1302.3307},
 primaryClass = {astro-ph.HE},
       adsurl = {https://ui.adsabs.harvard.edu/abs/2013Sci...339..807A}
}

@ARTICLE{HESS_SNR,
       author = {{H.~E.~S.~S. Collaboration} and {Abdalla}, H. and {Abramowski}, A. and {Aharonian}, F. and {Ait Benkhali}, F. and {Ang{\"u}ner}, E.~O. and {Arakawa}, M. and {Arrieta}, M. and {Aubert}, P. and {Backes}, M. and {Balzer}, A. and {Barnard}, M. and {Becherini}, Y. and {Becker Tjus}, J. and {Berge}, D. and {Bernhard}, S. and {Bernl{\"o}hr}, K. and {Blackwell}, R. and {B{\"o}ttcher}, M. and {Boisson}, C. and {Bolmont}, J. and {Bonnefoy}, S. and {Bordas}, P. and {Bregeon}, J. and {Brun}, F. and {Brun}, P. and {Bryan}, M. and {B{\"u}chele}, M. and {Bulik}, T. and {Capasso}, M. and {Caroff}, S. and {Carosi}, A. and {Casanova}, S. and {Cerruti}, M. and {Chakraborty}, N. and {Chaves}, R.~C.~G. and {Chen}, A. and {Chevalier}, J. and {Colafrancesco}, S. and {Condon}, B. and {Conrad}, J. and {Davids}, I.~D. and {Decock}, J. and {Deil}, C. and {Devin}, J. and {deWilt}, P. and {Dirson}, L. and {Djannati-Ata{\"\i}}, A. and {Donath}, A. and {Drury}, L.~O. 'C. and {Dutson}, K. and {Dyks}, J. and {Edwards}, T. and {Egberts}, K. and {Emery}, G. and {Ernenwein}, J. -P. and {Eschbach}, S. and {Farnier}, C. and {Fegan}, S. and {Fernandes}, M.~V. and {Fernandez}, D. and {Fiasson}, A. and {Fontaine}, G. and {Funk}, S. and {F{\"u}{\ss}ling}, M. and {Gabici}, S. and {Gallant}, Y.~A. and {Garrigoux}, T. and {Gat{\'e}}, F. and {Giavitto}, G. and {Giebels}, B. and {Glawion}, D. and {Glicenstein}, J.~F. and {Gottschall}, D. and {Grondin}, M. -H. and {Hahn}, J. and {Haupt}, M. and {Hawkes}, J. and {Heinzelmann}, G. and {Henri}, G. and {Hermann}, G. and {Hinton}, J.~A. and {Hofmann}, W. and {Hoischen}, C. and {Holch}, T.~L. and {Holler}, M. and {Horns}, D. and {Ivascenko}, A. and {Iwasaki}, H. and {Jacholkowska}, A. and {Jamrozy}, M. and {Jankowsky}, D. and {Jankowsky}, F. and {Jingo}, M. and {Jouvin}, L. and {Jung-Richardt}, I. and {Kastendieck}, M.~A. and {Katarzy{\'n}ski}, K. and {Katsuragawa}, M. and {Katz}, U. and {Kerszberg}, D. and {Khangulyan}, D. and {Kh{\'e}lifi}, B. and {King}, J. and {Klepser}, S. and {Klochkov}, D. and {Klu{\'z}niak}, W. and {Komin}, Nu. and {Kosack}, K. and {Krakau}, S. and {Kraus}, M. and {Kr{\"u}ger}, P.~P. and {Laffon}, H. and {Lamanna}, G. and {Lau}, J. and {Lees}, J. -P. and {Lefaucheur}, J. and {Lemi{\`e}re}, A. and {Lemoine-Goumard}, M. and {Lenain}, J. -P. and {Leser}, E. and {Lohse}, T. and {Lorentz}, M. and {Liu}, R. and {L{\'o}pez-Coto}, R. and {Lypova}, I. and {Malyshev}, D. and {Marandon}, V. and {Marcowith}, A. and {Mariaud}, C. and {Marx}, R. and {Maurin}, G. and {Maxted}, N. and {Mayer}, M. and {Meintjes}, P.~J. and {Meyer}, M. and {Mitchell}, A.~M.~W. and {Moderski}, R. and {Mohamed}, M. and {Mohrmann}, L. and {Mor{\r{a}}}, K. and {Moulin}, E. and {Murach}, T. and {Nakashima}, S. and {de Naurois}, M. and {Ndiyavala}, H. and {Niederwanger}, F. and {Niemiec}, J. and {Oakes}, L. and {O'Brien}, P. and {Odaka}, H. and {Ohm}, S. and {Ostrowski}, M. and {Oya}, I. and {Padovani}, M. and {Panter}, M. and {Parsons}, R.~D. and {Pekeur}, N.~W. and {Pelletier}, G. and {Perennes}, C. and {Petrucci}, P. -O. and {Peyaud}, B. and {Piel}, Q. and {Pita}, S. and {Poireau}, V. and {Poon}, H. and {Prokhorov}, D. and {Prokoph}, H. and {P{\"u}hlhofer}, G. and {Punch}, M. and {Quirrenbach}, A. and {Raab}, S. and {Rauth}, R. and {Reimer}, A. and {Reimer}, O. and {Renaud}, M. and {de los Reyes}, R. and {Rieger}, F. and {Rinchiuso}, L. and {Romoli}, C. and {Rowell}, G. and {Rudak}, B. and {Rulten}, C.~B. and {Safi-Harb}, S. and {Sahakian}, V. and {Saito}, S. and {Sanchez}, D.~A. and {Santangelo}, A. and {Sasaki}, M. and {Schlickeiser}, R. and {Sch{\"u}ssler}, F. and {Schulz}, A. and {Schwanke}, U. and {Schwemmer}, S. and {Seglar-Arroyo}, M. and {Settimo}, M. and {Seyffert}, A.~S. and {Shafi}, N. and {Shilon}, I. and {Shiningayamwe}, K. and {Simoni}, R. and {Sol}, H. and {Spanier}, F. and {Spir-Jacob}, M. and {Stawarz}, {\L}. and {Steenkamp}, R. and {Stegmann}, C. and {Steppa}, C. and {Sushch}, I. and {Takahashi}, T. and {Tavernet}, J. -P. and {Tavernier}, T. and {Taylor}, A.~M. and {Terrier}, R. and {Tibaldo}, L. and {Tiziani}, D. and {Tluczykont}, M. and {Trichard}, C. and {Tsirou}, M. and {Tsuji}, N. and {Tuffs}, R. and {Uchiyama}, Y. and {van der Walt}, D.~J. and {van Eldik}, C. and {van Rensburg}, C. and {van Soelen}, B. and {Vasileiadis}, G. and {Veh}, J. and {Venter}, C. and {Viana}, A. and {Vincent}, P. and {Vink}, J. and {Voisin}, F. and {V{\"o}lk}, H.~J. and {Vuillaume}, T. and {Wadiasingh}, Z. and {Wagner}, S.~J. and {Wagner}, P. and {Wagner}, R.~M. and {White}, R. and {Wierzcholska}, A. and {Willmann}, P. and {W{\"o}rnlein}, A. and {Wouters}, D. and {Yang}, R. and {Zaborov}, D. and {Zacharias}, M. and {Zanin}, R. and {Zdziarski}, A.~A. and {Zech}, A. and {Zefi}, F. and {Ziegler}, A. and {Zorn}, J. and {{\.Z}ywucka}, N.},
        title = "{Population study of Galactic supernova remnants at very high {\ensuremath{\gamma}}-ray energies with H.E.S.S.}",
      journal = {\aap},
         year = 2018,
        month = apr,
       volume = {612},
          eid = {A3},
        pages = {A3},
          doi = {10.1051/0004-6361/201732125},
archivePrefix = {arXiv},
       eprint = {1802.05172},
 primaryClass = {astro-ph.HE},
       adsurl = {https://ui.adsabs.harvard.edu/abs/2018A&A...612A...3H}
}

@ARTICLE{FICHTNER2024,
       author = {{Fichtner}, Yvonne A. and {Mackey}, Jonathan and {Grassitelli}, Luca and {Romano-D{\'\i}az}, Emilio and {Porciani}, Cristiano},
        title = "{Connecting stellar and galactic scales: Energetic feedback from stellar wind bubbles to supernova remnants}",
      journal = {\aap},
         year = 2024,
        month = oct,
       volume = {690},
          eid = {A72},
        pages = {A72},
          doi = {10.1051/0004-6361/202449638},
archivePrefix = {arXiv},
       eprint = {2402.11008},
 primaryClass = {astro-ph.GA},
       adsurl = {https://ui.adsabs.harvard.edu/abs/2024A&A...690A..72F}
}

@ARTICLE{DWARKADAS2005,
       author = {{Dwarkadas}, Vikram V.},
        title = "{The Evolution of Supernovae in Circumstellar Wind-Blown Bubbles. I. Introduction and One-Dimensional Calculations}",
      journal = {\apj},
         year = 2005,
        month = sep,
       volume = {630},
       number = {2},
        pages = {892-910},
          doi = {10.1086/432109},
archivePrefix = {arXiv},
       eprint = {astro-ph/0410464},
 primaryClass = {astro-ph},
       adsurl = {https://ui.adsabs.harvard.edu/abs/2005ApJ...630..892D}
}

@ARTICLE{DWARKADAS2007,
       author = {{Dwarkadas}, Vikram V.},
        title = "{The Evolution of Supernovae in Circumstellar Wind Bubbles. II. Case of a Wolf-Rayet Star}",
      journal = {\apj},
         year = 2007,
        month = sep,
       volume = {667},
       number = {1},
        pages = {226-247},
          doi = {10.1086/520670},
archivePrefix = {arXiv},
       eprint = {0706.1049},
 primaryClass = {astro-ph},
       adsurl = {https://ui.adsabs.harvard.edu/abs/2007ApJ...667..226D}
}

@ARTICLE{MEYER2024,
       author = {{Meyer}, D.~M. -A. and {Vel{\'a}zquez}, P.~F. and {Pohl}, M. and {Egberts}, K. and {Petrov}, M. and {Villagran}, M.~A. and {Torres}, D.~F. and {Batzofin}, R.},
        title = "{Supernova remnants of red supergiants: From barrels to loops}",
      journal = {\aap},
         year = 2024,
        month = jul,
       volume = {687},
          eid = {A127},
        pages = {A127},
          doi = {10.1051/0004-6361/202449706},
archivePrefix = {arXiv},
       eprint = {2404.07873},
 primaryClass = {astro-ph.HE},
       adsurl = {https://ui.adsabs.harvard.edu/abs/2024A&A...687A.127M}
}

@ARTICLE{Grunhut2017,
       author = {{Grunhut}, J.~H. and {Wade}, G.~A. and {Neiner}, C. and {Oksala}, M.~E. and {Petit}, V. and {Alecian}, E. and {Bohlender}, D.~A. and {Bouret}, J. -C. and {Henrichs}, H.~F. and {Hussain}, G.~A.~J. and {Kochukhov}, O. and {MiMeS Collaboration}},
        title = "{The MiMeS survey of Magnetism in Massive Stars: magnetic analysis of the O-type stars}",
      journal = {\mnras},
         year = 2017,
        month = feb,
       volume = {465},
       number = {2},
        pages = {2432-2470},
          doi = {10.1093/mnras/stw2743},
archivePrefix = {arXiv},
       eprint = {1610.07895},
 primaryClass = {astro-ph.SR},
       adsurl = {https://ui.adsabs.harvard.edu/abs/2017MNRAS.465.2432G}
}

@ARTICLE{Scholler2017,
       author = {{Sch{\"o}ller}, M. and {Hubrig}, S. and {Fossati}, L. and {Carroll}, T.~A. and {Briquet}, M. and {Oskinova}, L.~M. and {J{\"a}rvinen}, S. and {Ilyin}, I. and {Castro}, N. and {Morel}, T. and {Langer}, N. and {Przybilla}, N. and {Nieva}, M. -F. and {Kholtygin}, A.~F. and {Sana}, H. and {Herrero}, A. and {Barb{\'a}}, R.~H. and {de Koter}, A. and {BOB Collaboration}},
        title = "{B fields in OB stars (BOB): Concluding the FORS 2 observing campaign}",
      journal = {\aap},
         year = 2017,
        month = mar,
       volume = {599},
          eid = {A66},
        pages = {A66},
          doi = {10.1051/0004-6361/201628905},
archivePrefix = {arXiv},
       eprint = {1611.04502},
 primaryClass = {astro-ph.SR},
       adsurl = {https://ui.adsabs.harvard.edu/abs/2017A&A...599A..66S}
}

@ARTICLE{Frost2024,
       author = {{Frost}, A.~J. and {Sana}, H. and {Mahy}, L. and {Wade}, G. and {Barron}, J. and {Le Bouquin}, J. -B. and {M{\'e}rand}, A. and {Schneider}, F.~R.~N. and {Shenar}, T. and {Barb{\'a}}, R.~H. and {Bowman}, D.~M. and {Fabry}, M. and {Farhang}, A. and {Marchant}, P. and {Morrell}, N.~I. and {Smoker}, J.~V.},
        title = "{A magnetic massive star has experienced a stellar merger}",
      journal = {Science},
         year = 2024,
        month = apr,
       volume = {384},
       number = {6692},
        pages = {214-217},
          doi = {10.1126/science.adg7700},
archivePrefix = {arXiv},
       eprint = {2404.10167},
 primaryClass = {astro-ph.SR},
       adsurl = {https://ui.adsabs.harvard.edu/abs/2024Sci...384..214F}
}

@ARTICLE{Shenar2023,
       author = {{Shenar}, Tomer and {Wade}, Gregg A. and {Marchant}, Pablo and {Bagnulo}, Stefano and {Bodensteiner}, Julia and {Bowman}, Dominic M. and {Gilkis}, Avishai and {Langer}, Norbert and {Nicolas-Chen{\'e}}, Andr{\'e} and {Oskinova}, Lidia and {Van Reeth}, Timothy and {Sana}, Hugues and {St-Louis}, Nicole and {de Oliveira}, Alexandre Soares and {Todt}, Helge and {Toonen}, Silvia},
        title = "{A massive helium star with a sufficiently strong magnetic field to form a magnetar}",
      journal = {Science},
         year = 2023,
        month = aug,
       volume = {381},
       number = {6659},
        pages = {761-765},
          doi = {10.1126/science.ade3293},
archivePrefix = {arXiv},
       eprint = {2308.08591},
 primaryClass = {astro-ph.SR},
       adsurl = {https://ui.adsabs.harvard.edu/abs/2023Sci...381..761S}
}

@ARTICLE{Hubrig2020,
       author = {{Hubrig}, S. and {Sch{\"o}ller}, M. and {Cikota}, A. and {J{\"a}rvinen}, S.~P.},
        title = "{The search for magnetic fields in two Wolf-Rayet stars and the discovery of a variable magnetic field in WR 55}",
      journal = {\mnras},
         year = 2020,
        month = dec,
       volume = {499},
       number = {1},
        pages = {L116-L120},
          doi = {10.1093/mnrasl/slaa170},
archivePrefix = {arXiv},
       eprint = {2010.00983},
 primaryClass = {astro-ph.SR},
       adsurl = {https://ui.adsabs.harvard.edu/abs/2020MNRAS.499L.116H}
}

@ARTICLE{delaChevrotiere2014,
       author = {{de la Chevroti{\`e}re}, A. and {St-Louis}, N. and {Moffat}, A.~F.~J. and {MiMeS Collaboration}},
        title = "{Searching for Magnetic Fields in 11 Wolf-Rayet Stars: Analysis of Circular Polarization Measurements from ESPaDOnS}",
      journal = {\apj},
         year = 2014,
        month = feb,
       volume = {781},
       number = {2},
          eid = {73},
        pages = {73},
          doi = {10.1088/0004-637X/781/2/73},
       adsurl = {https://ui.adsabs.harvard.edu/abs/2014ApJ...781...73D}
}

@INCOLLECTION{Axford1972,
       author = {{Axford}, W.~I.},
        title = "{The Interaction of the Solar Wind With the Interstellar Medium}",
    booktitle = {NASA Special Publication},
         year = 1972,
       editor = {{Sonett}, Charles P. and {Coleman}, Paul Jerome and {Wilcox}, John Marsh},
       volume = {308},
        pages = {609},
       adsurl = {https://ui.adsabs.harvard.edu/abs/1972NASSP.308..609A}
}

@PHDTHESIS{Cranfill74,
  author = {{Cranfill}, Charles William},
        title = "{Flow Problems in Astrophysical Systems.}",
       school = {University of California, San Diego},
         year = 1974,
        month = sep,
       adsurl = {https://ui.adsabs.harvard.edu/abs/1974PhDT........73C}
}

@ARTICLE{TrueloveMcKee1999,
       author = {{Truelove}, J. Kelly and {McKee}, Christopher F.},
        title = "{Evolution of Nonradiative Supernova Remnants}",
      journal = {\apjs},
         year = 1999,
        month = feb,
       volume = {120},
       number = {2},
        pages = {299-326},
          doi = {10.1086/313176},
       adsurl = {https://ui.adsabs.harvard.edu/abs/1999ApJS..120..299T}
}

@ARTICLE{TenorioTagle1990,
       author = {{Tenorio-Tagle}, G. and {Rozyczka}, M. and {Bodenheimer}, P.},
        title = "{The hydrodynamics of superstructures produced by multi-supernova explosions}",
      journal = {\aap},
         year = 1990,
        month = oct,
       volume = {237},
       number = {1},
        pages = {207-214},
       adsurl = {https://ui.adsabs.harvard.edu/abs/1990A&A...237..207T}
}

@ARTICLE{Green2019,
       author = {{Green}, Samuel and {Mackey}, Jonathan and {Haworth}, Thomas J. and {Gvaramadze}, Vasilii V. and {Duffy}, Peter},
        title = "{Thermal emission from bow shocks. I. 2D hydrodynamic models of the Bubble Nebula}",
      journal = {\aap},
         year = 2019,
        month = may,
       volume = {625},
          eid = {A4},
        pages = {A4},
          doi = {10.1051/0004-6361/201834832},
archivePrefix = {arXiv},
       eprint = {1903.05505},
 primaryClass = {astro-ph.GA},
       adsurl = {https://ui.adsabs.harvard.edu/abs/2019A&A...625A...4G}
}

@ARTICLE{Wiersma2009,
       author = {{Wiersma}, Robert P.~C. and {Schaye}, Joop and {Smith}, Britton D.},
        title = "{The effect of photoionization on the cooling rates of enriched, astrophysical plasmas}",
      journal = {\mnras},
         year = 2009,
        month = feb,
       volume = {393},
       number = {1},
        pages = {99-107},
          doi = {10.1111/j.1365-2966.2008.14191.x},
archivePrefix = {arXiv},
       eprint = {0807.3748},
 primaryClass = {astro-ph},
       adsurl = {https://ui.adsabs.harvard.edu/abs/2009MNRAS.393...99W}
}

@ARTICLE{Henney2009,
       author = {{Henney}, William J. and {Arthur}, S. Jane and {de Colle}, Fabio and {Mellema}, Garrelt},
        title = "{Radiation-magnetohydrodynamic simulations of the photoionization of magnetized globules}",
      journal = {\mnras},
         year = 2009,
        month = sep,
       volume = {398},
       number = {1},
        pages = {157-175},
          doi = {10.1111/j.1365-2966.2009.15153.x},
archivePrefix = {arXiv},
       eprint = {0810.1531},
 primaryClass = {astro-ph},
       adsurl = {https://ui.adsabs.harvard.edu/abs/2009MNRAS.398..157H}
}

@ARTICLE{Hummer1994,
       author = {{Hummer}, D.~G.},
        title = "{Total Recombination and Energy Loss Coefficients for Hydrogenic Ions at Low Density for 10<T/E/Z/2<10/7K}",
      journal = {\mnras},
         year = 1994,
        month = may,
       volume = {268},
        pages = {109},
          doi = {10.1093/mnras/268.1.109},
       adsurl = {https://ui.adsabs.harvard.edu/abs/1994MNRAS.268..109H}
}

@BOOK{Rybicki1979,
       author = {{Rybicki}, George B. and {Lightman}, Alan P.},
        title = "{Radiative processes in astrophysics}",
         year = 1979,
        publisher = "John Wiley \& Sons, Inc.",
       adsurl = {https://ui.adsabs.harvard.edu/abs/1979rpa..book.....R}
}

@article{DEDNER2002,
title = {Hyperbolic Divergence Cleaning for the MHD Equations},
journal = {Journal of Computational Physics},
volume = {175},
number = {2},
pages = {645-673},
year = {2002},
issn = {0021-9991},
doi = {https://doi.org/10.1006/jcph.2001.6961},
url = {https://www.sciencedirect.com/science/article/pii/S002199910196961X},
author = {A. Dedner and F. Kemm and D. Kröner and C.-D. Munz and T. Schnitzer and M. Wesenberg}
}

@ARTICLE{Mackey2015,
       author = {{Mackey}, Jonathan and {Castro}, Norberto and {Fossati}, Luca and {Langer}, Norbert},
        title = "{Cold gas in hot star clusters: the wind from the red supergiant W26 in Westerlund 1}",
      journal = {\aap},
         year = 2015,
        month = oct,
       volume = {582},
          eid = {A24},
        pages = {A24},
          doi = {10.1051/0004-6361/201526159},
archivePrefix = {arXiv},
       eprint = {1508.07003},
 primaryClass = {astro-ph.GA},
       adsurl = {https://ui.adsabs.harvard.edu/abs/2015A&A...582A..24M}
}

@ARTICLE{Schneider2019,
       author = {{Schneider}, Fabian R.~N. and {Ohlmann}, Sebastian T. and {Podsiadlowski}, Philipp and {R{\"o}pke}, Friedrich K. and {Balbus}, Steven A. and {Pakmor}, R{\"u}diger and {Springel}, Volker},
        title = "{Stellar mergers as the origin of magnetic massive stars}",
      journal = {\nat},
         year = 2019,
        month = oct,
       volume = {574},
       number = {7777},
        pages = {211-214},
          doi = {10.1038/s41586-019-1621-5},
archivePrefix = {arXiv},
       eprint = {1910.14058},
 primaryClass = {astro-ph.SR},
       adsurl = {https://ui.adsabs.harvard.edu/abs/2019Natur.574..211S}
}

@ARTICLE{vanMarle2015,
       author = {{van Marle}, A.~J. and {Meliani}, Z. and {Marcowith}, A.},
        title = "{Shape and evolution of wind-blown bubbles of massive stars: on the effect of the interstellar magnetic field}",
      journal = {\aap},
         year = 2015,
        month = dec,
       volume = {584},
          eid = {A49},
        pages = {A49},
          doi = {10.1051/0004-6361/201425230},
archivePrefix = {arXiv},
       eprint = {1509.00192},
 primaryClass = {astro-ph.SR},
       adsurl = {https://ui.adsabs.harvard.edu/abs/2015A&A...584A..49V}
}

@ARTICLE{GarciaSegura1996b_LBV,
       author = {{Garcia-Segura}, G. and {Mac Low}, M. -M. and {Langer}, N.},
        title = "{The dynamical evolution of circumstellar gas around massive stars. I. The impact of the time sequence Ostar -> LBV -> WR star.}",
      journal = {\aap},
         year = 1996,
        month = jan,
       volume = {305},
        pages = {229},
       adsurl = {https://ui.adsabs.harvard.edu/abs/1996A&A...305..229G}
}

@ARTICLE{GarciaSegura1996a_RSG_WR,
       author = {{Garcia-Segura}, G. and {Langer}, N. and {Mac Low}, M. -M.},
        title = "{The hydrodynamic evolution of circumstellar gas around massive stars. II. The impact of the time sequence O star -> RSG -> WR star.}",
      journal = {\aap},
         year = 1996,
        month = dec,
       volume = {316},
        pages = {133-146},
       adsurl = {https://ui.adsabs.harvard.edu/abs/1996A&A...316..133G}
}

@ARTICLE{vanMarle2012,
       author = {{van Marle}, A.~J. and {Keppens}, R.},
        title = "{Multi-dimensional models of circumstellar shells around evolved massive stars}",
      journal = {\aap},
         year = 2012,
        month = nov,
       volume = {547},
          eid = {A3},
        pages = {A3},
          doi = {10.1051/0004-6361/201218957},
archivePrefix = {arXiv},
       eprint = {1209.4496},
 primaryClass = {astro-ph.SR},
       adsurl = {https://ui.adsabs.harvard.edu/abs/2012A&A...547A...3V}
}

@ARTICLE{Telezhinsky2012b,
       author = {{Telezhinsky}, I. and {Dwarkadas}, V.~V. and {Pohl}, M.},
        title = "{Time-dependent escape of cosmic rays from supernova remnants, and their interaction with dense media}",
      journal = {\aap},
         year = 2012,
        month = may,
       volume = {541},
          eid = {A153},
        pages = {A153},
          doi = {10.1051/0004-6361/201118639},
archivePrefix = {arXiv},
       eprint = {1112.3194},
 primaryClass = {astro-ph.HE},
       adsurl = {https://ui.adsabs.harvard.edu/abs/2012A&A...541A.153T}
}

@ARTICLE{Telezhinsky2013,
       author = {{Telezhinsky}, I. and {Dwarkadas}, V.~V. and {Pohl}, M.},
        title = "{Acceleration of cosmic rays by young core-collapse supernova remnants}",
      journal = {\aap},
         year = 2013,
        month = apr,
       volume = {552},
          eid = {A102},
        pages = {A102},
          doi = {10.1051/0004-6361/201220740},
archivePrefix = {arXiv},
       eprint = {1211.3627},
 primaryClass = {astro-ph.HE},
       adsurl = {https://ui.adsabs.harvard.edu/abs/2013A&A...552A.102T}
}

@ARTICLE{Langer2012,
       author = {{Langer}, N.},
        title = "{Presupernova Evolution of Massive Single and Binary Stars}",
      journal = {\araa},
         year = 2012,
        month = sep,
       volume = {50},
        pages = {107-164},
          doi = {10.1146/annurev-astro-081811-125534},
archivePrefix = {arXiv},
       eprint = {1206.5443},
 primaryClass = {astro-ph.SR},
       adsurl = {https://ui.adsabs.harvard.edu/abs/2012ARA&A..50..107L}
}

@ARTICLE{Dyson1972,
       author = {{Dyson}, J.~E. and {de Vries}, J.},
        title = "{The Dynamical Effects of Stellar Mass Loss on Diffuse Nebulae}",
      journal = {\aap},
         year = 1972,
        month = aug,
       volume = {20},
        pages = {223},
       adsurl = {https://ui.adsabs.harvard.edu/abs/1972A&A....20..223D}
}

@ARTICLE{Orlando2021,
       author = {{Orlando}, S. and {Wongwathanarat}, A. and {Janka}, H. -T. and {Miceli}, M. and {Ono}, M. and {Nagataki}, S. and {Bocchino}, F. and {Peres}, G.},
        title = "{The fully developed remnant of a neutrino-driven supernova. Evolution of ejecta structure and asymmetries in SNR Cassiopeia A}",
      journal = {\aap},
         year = 2021,
        month = jan,
       volume = {645},
          eid = {A66},
        pages = {A66},
          doi = {10.1051/0004-6361/202039335},
archivePrefix = {arXiv},
       eprint = {2009.01789},
 primaryClass = {astro-ph.HE},
       adsurl = {https://ui.adsabs.harvard.edu/abs/2021A&A...645A..66O}
}

@ARTICLE{Wongwathanarat2017,
       author = {{Wongwathanarat}, Annop and {Janka}, Hans-Thomas and {M{\"u}ller}, Ewald and {Pllumbi}, Else and {Wanajo}, Shinya},
        title = "{Production and Distribution of $^{44}$Ti and $^{56}$Ni in a Three-dimensional Supernova Model Resembling Cassiopeia A}",
      journal = {\apj},
         year = 2017,
        month = jun,
       volume = {842},
       number = {1},
          eid = {13},
        pages = {13},
          doi = {10.3847/1538-4357/aa72de},
archivePrefix = {arXiv},
       eprint = {1610.05643},
 primaryClass = {astro-ph.HE},
       adsurl = {https://ui.adsabs.harvard.edu/abs/2017ApJ...842...13W}
}

@ARTICLE{Boggs2015,
       author = {{Boggs}, S.~E. and {Harrison}, F.~A. and {Miyasaka}, H. and {Grefenstette}, B.~W. and {Zoglauer}, A. and {Fryer}, C.~L. and {Reynolds}, S.~P. and {Alexander}, D.~M. and {An}, H. and {Barret}, D. and {Christensen}, F.~E. and {Craig}, W.~W. and {Forster}, K. and {Giommi}, P. and {Hailey}, C.~J. and {Hornstrup}, A. and {Kitaguchi}, T. and {Koglin}, J.~E. and {Madsen}, K.~K. and {Mao}, P.~H. and {Mori}, K. and {Perri}, M. and {Pivovaroff}, M.~J. and {Puccetti}, S. and {Rana}, V. and {Stern}, D. and {Westergaard}, N.~J. and {Zhang}, W.~W.},
        title = "{$^{44}$Ti gamma-ray emission lines from SN1987A reveal an asymmetric explosion}",
      journal = {Science},
         year = 2015,
        month = may,
       volume = {348},
       number = {6235},
        pages = {670-671},
          doi = {10.1126/science.aaa2259},
       adsurl = {https://ui.adsabs.harvard.edu/abs/2015Sci...348..670B}
}

@ARTICLE{Wang2016,
       author = {{Wang}, Wei and {Li}, Zhuo},
        title = "{Hard X-Ray Emissions from Cassiopeia A Observed by INTEGRAL}",
      journal = {\apj},
         year = 2016,
        month = jul,
       volume = {825},
       number = {2},
          eid = {102},
        pages = {102},
          doi = {10.3847/0004-637X/825/2/102},
archivePrefix = {arXiv},
       eprint = {1605.00360},
 primaryClass = {astro-ph.HE},
       adsurl = {https://ui.adsabs.harvard.edu/abs/2016ApJ...825..102W}
}

@ARTICLE{Bell2008,
       author = {{Bell}, A.~R.},
        title = "{Cosmic ray acceleration by a supernova shock in a dense circumstellar plasma}",
      journal = {\mnras},
         year = 2008,
        month = apr,
       volume = {385},
       number = {4},
        pages = {1884-1892},
          doi = {10.1111/j.1365-2966.2008.13011.x},
       adsurl = {https://ui.adsabs.harvard.edu/abs/2008MNRAS.385.1884B}
}

@INPROCEEDINGS{Sushch2024,
       author = {{Sushch}, I. and {Le Roux}, J.~F. and {Brose}, R.},
        title = "{Particle acceleration at reflected shocks in supernova remnants}",
    booktitle = {38th International Cosmic Ray Conference},
         year = 2024,
        month = sep,
          eid = {262},
        pages = {262},
       adsurl = {https://ui.adsabs.harvard.edu/abs/2024icrc.confE.262S}
}

@ARTICLE{Zhekov2009,
       author = {{Zhekov}, Svetozar A. and {McCray}, Richard and {Dewey}, Daniel and {Canizares}, Claude R. and {Borkowski}, Kazimierz J. and {Burrows}, David N. and {Park}, Sangwook},
        title = "{High-Resolution X-Ray Spectroscopy of SNR 1987A: Chandra Letg and HETG Observations in 2007}",
      journal = {\apj},
         year = 2009,
        month = feb,
       volume = {692},
       number = {2},
        pages = {1190-1204},
          doi = {10.1088/0004-637X/692/2/1190},
archivePrefix = {arXiv},
       eprint = {0810.5313},
 primaryClass = {astro-ph},
       adsurl = {https://ui.adsabs.harvard.edu/abs/2009ApJ...692.1190Z}
}

@ARTICLE{Vink2022,
       author = {{Vink}, Jacco and {Patnaude}, Daniel J. and {Castro}, Daniel},
        title = "{The Forward and Reverse Shock Dynamics of Cassiopeia A}",
      journal = {\apj},
         year = 2022,
        month = apr,
       volume = {929},
       number = {1},
          eid = {57},
        pages = {57},
          doi = {10.3847/1538-4357/ac590f},
archivePrefix = {arXiv},
       eprint = {2201.08911},
 primaryClass = {astro-ph.HE},
       adsurl = {https://ui.adsabs.harvard.edu/abs/2022ApJ...929...57V}
}

@ARTICLE{Borkowski2018,
       author = {{Borkowski}, Kazimierz J. and {Reynolds}, Stephen P. and {Williams}, Brian J. and {Petre}, Robert},
        title = "{Expansion and Age of the X-Ray Synchrotron-dominated Supernova Remnant G330.2+1.0}",
      journal = {\apjl},
         year = 2018,
        month = dec,
       volume = {868},
       number = {2},
          eid = {L21},
        pages = {L21},
          doi = {10.3847/2041-8213/aaedb5},
archivePrefix = {arXiv},
       eprint = {1811.01998},
 primaryClass = {astro-ph.HE},
       adsurl = {https://ui.adsabs.harvard.edu/abs/2018ApJ...868L..21B}
}

@ARTICLE{Auriere2010,
       author = {{Auri{\`e}re}, M. and {Donati}, J. -F. and {Konstantinova-Antova}, R. and {Perrin}, G. and {Petit}, P. and {Roudier}, T.},
        title = "{The magnetic field of Betelgeuse: a local dynamo from giant convection cells?}",
      journal = {\aap},
         year = 2010,
        month = jun,
       volume = {516},
          eid = {L2},
        pages = {L2},
          doi = {10.1051/0004-6361/201014925},
archivePrefix = {arXiv},
       eprint = {1005.4845},
 primaryClass = {astro-ph.SR},
       adsurl = {https://ui.adsabs.harvard.edu/abs/2010A&A...516L...2A}
}

@ARTICLE{Tessore2017,
       author = {{Tessore}, B. and {L{\`e}bre}, A. and {Morin}, J. and {Mathias}, P. and {Josselin}, E. and {Auri{\`e}re}, M.},
        title = "{Measuring surface magnetic fields of red supergiant stars}",
      journal = {\aap},
         year = 2017,
        month = jul,
       volume = {603},
          eid = {A129},
        pages = {A129},
          doi = {10.1051/0004-6361/201730473},
archivePrefix = {arXiv},
       eprint = {1704.07761},
 primaryClass = {astro-ph.SR},
       adsurl = {https://ui.adsabs.harvard.edu/abs/2017A&A...603A.129T}
}

@ARTICLE{Varma2021,
       author = {{Varma}, Vishnu and {M{\"u}ller}, Bernhard},
        title = "{3D simulations of oxygen shell burning with and without magnetic fields}",
      journal = {\mnras},
         year = 2021,
        month = jun,
       volume = {504},
       number = {1},
        pages = {636-647},
          doi = {10.1093/mnras/stab883},
archivePrefix = {arXiv},
       eprint = {2101.00213},
 primaryClass = {astro-ph.SR},
       adsurl = {https://ui.adsabs.harvard.edu/abs/2021MNRAS.504..636V}
}

@ARTICLE{GreMac21,
author = {{Green}, Samuel and {Mackey}, Jonathan},
title = "{PyPion: Post-processing code for PION simulation data}",
journal = {Astrophysics Source Code Library},
year = 2021,
month = mar,
archivePrefix = "ascl",
eprint = {2103.026},
adsurl = {https://ui.adsabs.harvard.edu/abs/2021ascl.soft03026G}
}

@ARTICLE{astropy:2018,
       author = {{Astropy Collaboration} and {Price-Whelan}, A.~M. and
         {Sip{H{o}}cz}, B.~M. and {G{"u}nther}, H.~M. and {Lim}, P.~L. and
         {Crawford}, S.~M. and {Conseil}, S. and {Shupe}, D.~L. and
         {Craig}, M.~W. and {Dencheva}, N. and {Ginsburg}, A. and {Vand
        erPlas}, J.~T. and {Bradley}, L.~D. and {P{'e}rez-Su{'a}rez}, D. and
         {de Val-Borro}, M. and {Aldcroft}, T.~L. and {Cruz}, K.~L. and
         {Robitaille}, T.~P. and {Tollerud}, E.~J. and {Ardelean}, C. and
         {Babej}, T. and {Bach}, Y.~P. and {Bachetti}, M. and {Bakanov}, A.~V. and
         {Bamford}, S.~P. and {Barentsen}, G. and {Barmby}, P. and
         {Baumbach}, A. and {Berry}, K.~L. and {Biscani}, F. and {Boquien}, M. and
         {Bostroem}, K.~A. and {Bouma}, L.~G. and {Brammer}, G.~B. and
         {Bray}, E.~M. and {Breytenbach}, H. and {Buddelmeijer}, H. and
         {Burke}, D.~J. and {Calderone}, G. and {Cano Rodr{'i}guez}, J.~L. and
         {Cara}, M. and {Cardoso}, J.~V.~M. and {Cheedella}, S. and {Copin}, Y. and
         {Corrales}, L. and {Crichton}, D. and {D'Avella}, D. and {Deil}, C. and
         {Depagne}, {'E}. and {Dietrich}, J.~P. and {Donath}, A. and
         {Droettboom}, M. and {Earl}, N. and {Erben}, T. and {Fabbro}, S. and
         {Ferreira}, L.~A. and {Finethy}, T. and {Fox}, R.~T. and
         {Garrison}, L.~H. and {Gibbons}, S.~L.~J. and {Goldstein}, D.~A. and
         {Gommers}, R. and {Greco}, J.~P. and {Greenfield}, P. and
         {Groener}, A.~M. and {Grollier}, F. and {Hagen}, A. and {Hirst}, P. and
         {Homeier}, D. and {Horton}, A.~J. and {Hosseinzadeh}, G. and {Hu}, L. and
         {Hunkeler}, J.~S. and {Ivezi{'c}}, {{Z}}. and {Jain}, A. and
         {Jenness}, T. and {Kanarek}, G. and {Kendrew}, S. and {Kern}, N.~S. and
         {Kerzendorf}, W.~E. and {Khvalko}, A. and {King}, J. and {Kirkby}, D. and
         {Kulkarni}, A.~M. and {Kumar}, A. and {Lee}, A. and {Lenz}, D. and
         {Littlefair}, S.~P. and {Ma}, Z. and {Macleod}, D.~M. and
         {Mastropietro}, M. and {McCully}, C. and {Montagnac}, S. and
         {Morris}, B.~M. and {Mueller}, M. and {Mumford}, S.~J. and {Muna}, D. and
         {Murphy}, N.~A. and {Nelson}, S. and {Nguyen}, G.~H. and
         {Ninan}, J.~P. and {N{"o}the}, M. and {Ogaz}, S. and {Oh}, S. and
         {Parejko}, J.~K. and {Parley}, N. and {Pascual}, S. and {Patil}, R. and
         {Patil}, A.~A. and {Plunkett}, A.~L. and {Prochaska}, J.~X. and
         {Rastogi}, T. and {Reddy Janga}, V. and {Sabater}, J. and
         {Sakurikar}, P. and {Seifert}, M. and {Sherbert}, L.~E. and
         {Sherwood-Taylor}, H. and {Shih}, A.~Y. and {Sick}, J. and
         {Silbiger}, M.~T. and {Singanamalla}, S. and {Singer}, L.~P. and
         {Sladen}, P.~H. and {Sooley}, K.~A. and {Sornarajah}, S. and
         {Streicher}, O. and {Teuben}, P. and {Thomas}, S.~W. and
         {Tremblay}, G.~R. and {Turner}, J.~E.~H. and {Terr{'o}n}, V. and
         {van Kerkwijk}, M.~H. and {de la Vega}, A. and {Watkins}, L.~L. and
         {Weaver}, B.~A. and {Whitmore}, J.~B. and {Woillez}, J. and
         {Zabalza}, V. and {Astropy Contributors}},
        title = "{The Astropy Project: Building an Open-science Project and Status of the v2.0 Core Package}",
      journal = {\aj},
         year = 2018,
        month = sep,
       volume = {156},
       number = {3},
          eid = {123},
        pages = {123},
          doi = {10.3847/1538-3881/aabc4f},
archivePrefix = {arXiv},
       eprint = {1801.02634},
 primaryClass = {astro-ph.IM},
       adsurl = {https://ui.adsabs.harvard.edu/abs/2018AJ....156..123A}
}

@Article{HarMilVan20,
 title         = {Array programming with {NumPy}},
 author        = {Charles R. Harris and K. Jarrod Millman and St{'{e}}fan J.
                 van der Walt and Ralf Gommers and Pauli Virtanen and David
                 Cournapeau and Eric Wieser and Julian Taylor and Sebastian
                 Berg and Nathaniel J. Smith and Robert Kern and Matti Picus
                 and Stephan Hoyer and Marten H. van Kerkwijk and Matthew
                 Brett and Allan Haldane and Jaime Fern{'{a}}ndez del
                 R{'{\i}}o and Mark Wiebe and Pearu Peterson and Pierre
                 G{'{e}}rard-Marchant and Kevin Sheppard and Tyler Reddy and
                 Warren Weckesser and Hameer Abbasi and Christoph Gohlke and
                 Travis E. Oliphant},
 year          = {2020},
 month         = sep,
 journal       = {Nature},
 volume        = {585},
 number        = {7825},
 pages         = {357--362},
 doi           = {10.1038/s41586-020-2649-2},
 publisher     = {Springer Science and Business Media {LLC}},
 url           = {https://doi.org/10.1038/s41586-020-2649-2}
}

@Article{Hun07,
  Author    = {Hunter, J. D.},
  Title     = {Matplotlib: A 2D graphics environment},
  Journal   = {Computing in Science \& Engineering},
  Volume    = {9},
  Number    = {3},
  Pages     = {90--95},
  publisher = {IEEE COMPUTER SOC},
  doi       = {10.1109/MCSE.2007.55},
  year      = 2007
}

@ARTICLE{TurSmiOis11,
   author = {{Turk}, M.~J. and {Smith}, B.~D. and {Oishi}, J.~S. and {Skory}, S. and
     {Skillman}, S.~W. and {Abel}, T. and {Norman}, M.~L.},
    title = "{yt: A Multi-code Analysis Toolkit for Astrophysical Simulation Data}",
  journal = {The Astrophysical Journal Supplement Series},
archivePrefix = "arXiv",
   eprint = {1011.3514},
 primaryClass = "astro-ph.IM",
     year = 2011,
    month = jan,
   volume = 192,
      eid = {9},
    pages = {9},
      doi = {10.1088/0067-0049/192/1/9},
   adsurl = {https://ui.adsabs.harvard.edu/abs/2011ApJS..192....9T}
}

@ARTICLE{Meyer2017,
       author = {{Meyer}, D.~M. -A. and {Mignone}, A. and {Kuiper}, R. and {Raga}, A.~C. and {Kley}, W.},
        title = "{Bow shock nebulae of hot massive stars in a magnetized medium}",
      journal = {\mnras},
         year = 2017,
        month = jan,
       volume = {464},
       number = {3},
        pages = {3229-3248},
          doi = {10.1093/mnras/stw2537},
archivePrefix = {arXiv},
       eprint = {1610.00543},
 primaryClass = {astro-ph.SR},
       adsurl = {https://ui.adsabs.harvard.edu/abs/2017MNRAS.464.3229M}
}

@ARTICLE{Gvaramadze2012,
       author = {{Gvaramadze}, V.~V. and {Weidner}, C. and {Kroupa}, P. and {Pflamm-Altenburg}, J.},
        title = "{Field O stars: formed in situ or as runaways?}",
      journal = {\mnras},
         year = 2012,
        month = aug,
       volume = {424},
       number = {4},
        pages = {3037-3049},
          doi = {10.1111/j.1365-2966.2012.21452.x},
archivePrefix = {arXiv},
       eprint = {1206.1596},
 primaryClass = {astro-ph.SR},
       adsurl = {https://ui.adsabs.harvard.edu/abs/2012MNRAS.424.3037G}
}

@ARTICLE{Oey2013,
       author = {{Oey}, M.~S. and {Lamb}, J.~B. and {Kushner}, C.~T. and {Pellegrini}, E.~W. and {Graus}, A.~S.},
        title = "{A Sample of OB Stars that Formed in the Field}",
      journal = {\apj},
         year = 2013,
        month = may,
       volume = {768},
       number = {1},
          eid = {66},
        pages = {66},
          doi = {10.1088/0004-637X/768/1/66},
archivePrefix = {arXiv},
       eprint = {1303.1550},
 primaryClass = {astro-ph.GA},
       adsurl = {https://ui.adsabs.harvard.edu/abs/2013ApJ...768...66O}
}

@ARTICLE{Guarcello2024,
       author = {{Guarcello}, M.~G. and {Almendros-Abad}, V. and {Lovell}, J.~B. and {Monsch}, K. and {Mu{\v{z}}i{\'c}}, K. and {Mart{\'\i}nez-Galarza}, J.~R. and {Drake}, J.~J. and {Anastasopoulou}, K. and {Andersen}, M. and {Argiroffi}, C. and et al.},
        title = "{EWOCS-III: JWST observations of the supermassive star cluster Westerlund 1}",
      journal = {\aap},
         year = 2025,
        month = jan,
       volume = {693},
          eid = {A120},
        pages = {A120},
          doi = {10.1051/0004-6361/202452150},
archivePrefix = {arXiv},
       eprint = {2411.13051},
 primaryClass = {astro-ph.SR},
       adsurl = {https://ui.adsabs.harvard.edu/abs/2025A&A...693A.120G}
}

@ARTICLE{Fesen2025,
       author = {{Fesen}, Robert A. and {Milisavljevic}, Dan and {Patnaude}, Daniel and {Chevalier}, Roger A. and {Raymond}, John C. and {Brumback}, McKinley and {Weil}, Kathryn E.},
        title = "{Cassiopeia A's Reverse Shock and Its Effects on the Expanding SN Ejecta}",
      journal = {\apjs},
         year = 2025,
        month = may,
       volume = {278},
       number = {1},
          eid = {17},
        pages = {17},
          doi = {10.3847/1538-4365/adbf15},
archivePrefix = {arXiv},
       eprint = {2501.07708},
 primaryClass = {astro-ph.HE},
       adsurl = {https://ui.adsabs.harvard.edu/abs/2025ApJS..278...17F}
}

@ARTICLE{Das2024,
       author = {{Das}, Samata and {Brose}, Robert and {Pohl}, Martin and {Meyer}, Dominique M. -A. and {Sushch}, Iurii},
        title = "{Particle acceleration, escape, and non-thermal emission from core-collapse supernovae inside non-identical wind-blown bubbles}",
      journal = {\aap},
         year = 2024,
        month = sep,
       volume = {689},
          eid = {A9},
        pages = {A9},
          doi = {10.1051/0004-6361/202245680},
archivePrefix = {arXiv},
       eprint = {2408.15839},
 primaryClass = {astro-ph.HE},
       adsurl = {https://ui.adsabs.harvard.edu/abs/2024A&A...689A...9D}
}

@ARTICLE{Vink2001,
       author = {{Vink}, Jorick S. and {de Koter}, A. and {Lamers}, H.~J.~G.~L.~M.},
        title = "{Mass-loss predictions for O and B stars as a function of metallicity}",
      journal = {\aap},
         year = 2001,
        month = apr,
       volume = {369},
        pages = {574-588},
          doi = {10.1051/0004-6361:20010127},
archivePrefix = {arXiv},
       eprint = {astro-ph/0101509},
 primaryClass = {astro-ph},
       adsurl = {https://ui.adsabs.harvard.edu/abs/2001A&A...369..574V}
}

@ARTICLE{Nieuwenhuijzen1990,
       author = {{Nieuwenhuijzen}, H. and {de Jager}, C.},
        title = "{Parametrization of stellar rates of mass loss as functions of the fundamental stellar parameters M, L, and R.}",
      journal = {\aap},
         year = 1990,
        month = may,
       volume = {231},
        pages = {134-136},
       adsurl = {https://ui.adsabs.harvard.edu/abs/1990A&A...231..134N}
}

@ARTICLE{Nugis2000,
       author = {{Nugis}, T. and {Lamers}, H.~J.~G.~L.~M.},
        title = "{Mass-loss rates of Wolf-Rayet stars as a function of stellar parameters}",
      journal = {\aap},
         year = 2000,
        month = aug,
       volume = {360},
        pages = {227-244},
       adsurl = {https://ui.adsabs.harvard.edu/abs/2000A&A...360..227N}
}

@ARTICLE{Aguilera-Dena2023,
       author = {{Aguilera-Dena}, David R. and {M{\"u}ller}, Bernhard and {Antoniadis}, John and {Langer}, Norbert and {Dessart}, Luc and {Vigna-G{\'o}mez}, Alejandro and {Yoon}, Sung-Chul},
        title = "{Stripped-envelope stars in different metallicity environments. II. Type I supernovae and compact remnants}",
      journal = {\aap},
         year = 2023,
        month = mar,
       volume = {671},
          eid = {A134},
        pages = {A134},
          doi = {10.1051/0004-6361/202243519},
archivePrefix = {arXiv},
       eprint = {2204.00025},
 primaryClass = {astro-ph.SR},
       adsurl = {https://ui.adsabs.harvard.edu/abs/2023A&A...671A.134A}
}

@ARTICLE{Owocki2004,
       author = {{Owocki}, Stanley P. and {ud-Doula}, Asif},
        title = "{The Effect of Magnetic Field Tilt and Divergence on the Mass Flux and Flow Speed in a Line-driven Stellar Wind}",
      journal = {\apj},
         year = 2004,
        month = jan,
       volume = {600},
       number = {2},
        pages = {1004-1015},
          doi = {10.1086/380123},
archivePrefix = {arXiv},
       eprint = {astro-ph/0310176},
 primaryClass = {astro-ph},
       adsurl = {https://ui.adsabs.harvard.edu/abs/2004ApJ...600.1004O}
}

@ARTICLE{Oskinova2013,
       author = {{Oskinova}, L.~M. and {Steinke}, M. and {Hamann}, W. -R. and {Sander}, A. and {Todt}, H. and {Liermann}, A.},
        title = "{One of the most massive stars in the Galaxy may have formed in isolation}",
      journal = {\mnras},
         year = 2013,
        month = dec,
       volume = {436},
       number = {4},
        pages = {3357-3365},
          doi = {10.1093/mnras/stt1817},
archivePrefix = {arXiv},
       eprint = {1309.7651},
 primaryClass = {astro-ph.SR},
       adsurl = {https://ui.adsabs.harvard.edu/abs/2013MNRAS.436.3357O}
}

@ARTICLE{Vargas-Salazar2020,
       author = {{Vargas-Salazar}, Irene and {Oey}, M.~S. and {Barnes}, Jesse R. and {Chen}, Xinyi and {Castro}, N. and {Kratter}, Kaitlin M. and {Faerber}, Timothy A.},
        title = "{A Search for In Situ Field OB Star Formation in the Small Magellanic Cloud}",
      journal = {\apj},
         year = 2020,
        month = nov,
       volume = {903},
       number = {1},
          eid = {42},
        pages = {42},
          doi = {10.3847/1538-4357/abbb95},
archivePrefix = {arXiv},
       eprint = {2009.12379},
 primaryClass = {astro-ph.SR},
       adsurl = {https://ui.adsabs.harvard.edu/abs/2020ApJ...903...42V}
}

@ARTICLE{Jin2025,
       author = {{Jin}, Harim and {Langer}, Norbert and {Ercolino}, Andrea and {de Mink}, Selma E.},
        title = "{A comprehensive grid of massive binary evolution models for the Galaxy - Surface properties of post-mass transfer stars}",
      journal = {arXiv e-prints},
         year = 2025,
        month = oct,
          eid = {arXiv:2510.19965},
        pages = {arXiv:2510.19965},
          doi = {10.48550/arXiv.2510.19965},
archivePrefix = {arXiv},
       eprint = {2510.19965},
 primaryClass = {astro-ph.SR},
       adsurl = {https://ui.adsabs.harvard.edu/abs/2025arXiv251019965J}
}

@ARTICLE{Yoon2017,
       author = {{Yoon}, Sung-Chul},
        title = "{Towards a better understanding of the evolution of Wolf-Rayet stars and Type Ib/Ic supernova progenitors}",
      journal = {\mnras},
         year = 2017,
        month = oct,
       volume = {470},
       number = {4},
        pages = {3970-3980},
          doi = {10.1093/mnras/stx1496},
archivePrefix = {arXiv},
       eprint = {1706.04716},
 primaryClass = {astro-ph.SR},
       adsurl = {https://ui.adsabs.harvard.edu/abs/2017MNRAS.470.3970Y}
}

@BOOK{Shu1992,
       author = {{Shu}, F.~H.},
        title = "{The physics of astrophysics. Volume II: Gas dynamics.}",
         year = 1992,
       adsurl = {https://ui.adsabs.harvard.edu/abs/1992pavi.book.....S}
}

@ARTICLE{vanVeelen2009,
       author = {{van Veelen}, B. and {Langer}, N. and {Vink}, J. and {Garc{\'\i}a-Segura}, G. and {van Marle}, A.~J.},
        title = "{The hydrodynamics of the supernova remnant Cassiopeia A. The influence of the progenitor evolution on the velocity structure and clumping}",
      journal = {\aap},
         year = 2009,
        month = aug,
       volume = {503},
       number = {2},
        pages = {495-503},
          doi = {10.1051/0004-6361/200912393},
archivePrefix = {arXiv},
       eprint = {0907.1197},
 primaryClass = {astro-ph.HE},
       adsurl = {https://ui.adsabs.harvard.edu/abs/2009A&A...503..495V}
}

@ARTICLE{Meyer2015,
       author = {{Meyer}, D.~M. -A. and {Langer}, N. and {Mackey}, J. and {Vel{\'a}zquez}, P.~F. and {Gusdorf}, A.},
        title = "{Asymmetric supernova remnants generated by Galactic, massive runaway stars}",
      journal = {\mnras},
         year = 2015,
        month = jul,
       volume = {450},
       number = {3},
        pages = {3080-3100},
          doi = {10.1093/mnras/stv898},
archivePrefix = {arXiv},
       eprint = {1508.03347},
 primaryClass = {astro-ph.HE},
       adsurl = {https://ui.adsabs.harvard.edu/abs/2015MNRAS.450.3080M}
}

@ARTICLE{Ercolino2024a,
       author = {{Ercolino}, A. and {Jin}, H. and {Langer}, N. and {Dessart}, L.},
        title = "{Interacting supernovae from wide massive binary systems}",
      journal = {\aap},
         year = 2024,
        month = may,
       volume = {685},
          eid = {A58},
        pages = {A58},
          doi = {10.1051/0004-6361/202347646},
archivePrefix = {arXiv},
       eprint = {2308.01819},
 primaryClass = {astro-ph.SR},
       adsurl = {https://ui.adsabs.harvard.edu/abs/2024A&A...685A..58E}
}

@ARTICLE{Ercolino2024b,
       author = {{Ercolino}, A. and {Jin}, H. and {Langer}, N. and {Dessart}, L.},
        title = "{Mass-transferring binary stars as progenitors of interacting hydrogen-free supernovae}",
      journal = {\aap},
         year = 2025,
        month = apr,
       volume = {696},
          eid = {A103},
        pages = {A103},
          doi = {10.1051/0004-6361/202453426},
archivePrefix = {arXiv},
       eprint = {2412.09893},
 primaryClass = {astro-ph.SR},
       adsurl = {https://ui.adsabs.harvard.edu/abs/2025A&A...696A.103E}
}

@ARTICLE{Bell04,
       author = {{Bell}, A.~R.},
        title = "{Turbulent amplification of magnetic field and diffusive shock acceleration of cosmic rays}",
      journal = {\mnras},
         year = 2004,
        month = sep,
       volume = {353},
       number = {2},
        pages = {550-558},
          doi = {10.1111/j.1365-2966.2004.08097.x},
       adsurl = {https://ui.adsabs.harvard.edu/abs/2004MNRAS.353..550B}
}

@BOOK{GinzburgBook,
       author = {{Ginzburg}, V.~L. and {Syrovatskii}, S.~I.},
        title = "{The Origin of Cosmic Rays}",
        publisher = "Pergamon",
         year = 1964,
       adsurl = {https://ui.adsabs.harvard.edu/abs/1964ocr..book.....G}
}

@ARTICLE{2022Sci...376...77H,
       author = {{H.~E.~S.~S. Collaboration} and {Aharonian}, F. and {Ait Benkhali}, F. and {Ang{\"u}ner}, E.~O. and {Ashkar}, H. and {Backes}, M. and {Baghmanyan}, V. and {Barbosa Martins}, V. and {Batzofin}, R. and {Becherini}, Y. and {Berge}, D. and {Bernl{\"o}hr}, K. and {Bi}, B. and {B{\"o}ttcher}, M. and {Boisson}, C. and {Bolmont}, J. and {de Bony de Lavergne}, M. and {Breuhaus}, M. and {Brose}, R. and {Brun}, F. and {Caroff}, S. and {Casanova}, S. and {Cerruti}, M. and {Chand}, T. and {Chen}, A. and {Cotter}, G. and {Damascene Mbarubucyeye}, J. and {Djannati-Ata{\"\i}}, A. and {Dmytriiev}, A. and {Doroshenko}, V. and {Duffy}, C. and {Egberts}, K. and {Ernenwein}, J. -P. and {Fegan}, S. and {Feijen}, K. and {Fiasson}, A. and {Fichet de Clairfontaine}, G. and {Fontaine}, G. and {F{\"u}{\ss}ling}, M. and {Funk}, S. and {Gabici}, S. and {Gallant}, Y.~A. and {Ghafourizadeh}, S. and {Giavitto}, G. and {Giunti}, L. and {Glawion}, D. and {Glicenstein}, J.~F. and {Grondin}, M. -H. and {Hermann}, G. and {Hinton}, J.~A. and {H{\"o}rbe}, M. and {Hofmann}, W. and {Hoischen}, C. and {Holch}, T.~L. and {Holler}, M. and {Horns}, D. and {Huang}, Zhiqiu and {Jamrozy}, M. and {Jankowsky}, F. and {Jung-Richardt}, I. and {Kasai}, E. and {Katarzy{\'n}ski}, K. and {Katz}, U. and {Khangulyan}, D. and {Kh{\'e}lifi}, B. and {Klepser}, S. and {Klu{\'z}niak}, W. and {Komin}, Nu. and {Konno}, R. and {Kosack}, K. and {Kostunin}, D. and {Le Stum}, S. and {Lemi{\`e}re}, A. and {Lemoine-Goumard}, M. and {Lenain}, J. -P. and {Leuschner}, F. and {Lohse}, T. and {Luashvili}, A. and {Lypova}, I. and {Mackey}, J. and {Malyshev}, D. and {Malyshev}, D. and {Marandon}, V. and {Marchegiani}, P. and {Marcowith}, A. and {Mart{\'\i}-Devesa}, G. and {Marx}, R. and {Maurin}, G. and {Meyer}, M. and {Mitchell}, A. and {Moderski}, R. and {Mohrmann}, L. and {Montanari}, A. and {Moulin}, E. and {Muller}, J. and {Murach}, T. and {Nakashima}, K. and {de Naurois}, M. and {Nayerhoda}, A. and {Niemiec}, J. and {Priyana Noel}, A. and {O{\textquoteright}Brien}, P. and {Ohm}, S. and {Olivera-Nieto}, L. and {de Ona Wilhelmi}, E. and {Ostrowski}, M. and {Panny}, S. and {Panter}, M. and {Parsons}, R.~D. and {Peron}, G. and {Pita}, S. and {Poireau}, V. and {Prokhorov}, D.~A. and {Prokoph}, H. and {P{\"u}hlhofer}, G. and {Punch}, M. and {Quirrenbach}, A. and {Reichherzer}, P. and {Reimer}, A. and {Reimer}, O. and {Renaud}, M. and {Reville}, B. and {Rieger}, F. and {Rowell}, G. and {Rudak}, B. and {Rueda Ricarte}, H. and {Ruiz-Velasco}, E. and {Sahakian}, V. and {Sailer}, S. and {Salzmann}, H. and {Sanchez}, D.~A. and {Santangelo}, A. and {Sasaki}, M. and {Sch{\"a}fer}, J. and {Sch{\"u}ssler}, F. and {Schutte}, H.~M. and {Schwanke}, U. and {Senniappan}, M. and {Shapopi}, J.~N.~S. and {Simoni}, R. and {Sinha}, A. and {Sol}, H. and {Specovius}, A. and {Spencer}, S. and {Stawarz}, {\L}. and {Steinmassl}, S. and {Steppa}, C. and {Takahashi}, T. and {Tanaka}, T. and {Taylor}, A.~M. and {Terrier}, R. and {Thorpe-Morgan}, C. and {Tsirou}, M. and {Tsuji}, N. and {Tuffs}, R. and {Uchiyama}, Y. and {Unbehaun}, T. and {van Eldik}, C. and {van Soelen}, B. and {Veh}, J. and {Venter}, C. and {Vink}, J. and {Wagner}, S.~J. and {Werner}, F. and {White}, R. and {Wierzcholska}, A. and {Wong}, Yu Wun and {Yusafzai}, A. and {Zacharias}, M. and {Zargaryan}, D. and {Zdziarski}, A.~A. and {Zech}, A. and {Zhu}, S.~J. and {Zouari}, S. and {{\.Z}ywucka}, N.},
        title = "{Time-resolved hadronic particle acceleration in the recurrent nova RS Ophiuchi}",
      journal = {Science},
         year = 2022,
        month = apr,
       volume = {376},
       number = {6588},
        pages = {77-80},
          doi = {10.1126/science.abn0567},
archivePrefix = {arXiv},
       eprint = {2202.08201},
 primaryClass = {astro-ph.HE},
       adsurl = {https://ui.adsabs.harvard.edu/abs/2022Sci...376...77H}
}

@ARTICLE{Larkin25,
	author = {{Larkin}, C. ~J. ~K. and {Mackey}, J. and {Haworth, T. J.} and {Sander, A. A. C.}},
	title = {Investigating dusty red supergiant outflows in Westerlund 1 with 3D hydrodynamic simulations},
	DOI= "10.1051/0004-6361/202554334",
	url= "https://doi.org/10.1051/0004-6361/202554334",
	journal = {A\&A},
	year = 2025,
	volume = 700,
	pages = "A60",
}

@ARTICLE{Orlando2025GM,
       author = {{Orlando}, S. and {Janka}, H. -T. and {Wongwathanarat}, A. and {Bocchino}, F. and {De Looze}, I. and {Milisavljevic}, D. and {Miceli}, M. and {Temim}, T. and {Rho}, J. and {Nagataki}, S. and {Ono}, M. and {Sapienza}, V. and {Greco}, E.},
        title = "{Origin of holes and rings in the Green Monster of Cassiopeia A: Insights from 3D magnetohydrodynamic simulations}",
      journal = {\aap},
         year = 2025,
        month = apr,
       volume = {696},
          eid = {A188},
        pages = {A188},
          doi = {10.1051/0004-6361/202553902},
archivePrefix = {arXiv},
       eprint = {2503.14455},
 primaryClass = {astro-ph.HE},
       adsurl = {https://ui.adsabs.harvard.edu/abs/2025A&A...696A.188O}
}

@ARTICLE{Orlando20251987,
       author = {{Orlando}, S. and {Miceli}, M. and {Ono}, M. and {Nagataki}, S. and {Aloy}, M. -A. and {Bocchino}, F. and {Gabler}, M. and {Giudici}, B. and {Giuffrida}, R. and {Greco}, E. and {La Malfa}, G. and {Lee}, S. -H. and {Obergaulinger}, M. and {Petruk}, O. and {Sapienza}, V. and {Ustamujic}, S. and {Weng}, J.},
        title = "{Tracing the ejecta structure of supernova 1987A: Insights and diagnostics from 3D magnetohydrodynamic simulations}",
      journal = {\aap},
         year = 2025,
        month = jul,
       volume = {699},
          eid = {A305},
        pages = {A305},
          doi = {10.1051/0004-6361/202554862},
archivePrefix = {arXiv},
       eprint = {2504.19896},
 primaryClass = {astro-ph.HE},
       adsurl = {https://ui.adsabs.harvard.edu/abs/2025A&A...699A.305O}
}

@BOOK{LamersCassinelli1999,
       author = {{Lamers}, Henny J.~G.~L.~M. and {Cassinelli}, Joseph P.},
        title = "{Introduction to Stellar Winds}",
         year = 1999,
       adsurl = {https://ui.adsabs.harvard.edu/abs/1999isw..book.....L}
}

\label{LastPage}
\end{document}